\documentclass[a4paper,12pt]{article}
\usepackage{amsfonts}
\usepackage{latexsym}
\usepackage{amsmath}
\usepackage{amssymb}
\usepackage{amssymb}
\usepackage{slashed}
\usepackage{upgreek}
\usepackage{mathrsfs}
\usepackage{bbm}

\usepackage{relsize}
\usepackage{graphicx}

\newcommand{\newsection}{    
\setcounter{equation}{0}\section}
\def\appendix#1{\addtocounter{section}{1}\setcounter{equation}{0}
\renewcommand{\thesection}{\Alph{section}}
\section*{Appendix \thesection\protect\indent \parbox[t]{11.15cm}{#1}}
\addcontentsline{toc}{section}{Appendix \thesection\ \ \ #1}}

\def\bbe{{\bf{e}}}
\font\mybb=msbm10 at 11pt

\def\bb#1{\hbox{\mybb#1}}

\def\bR {\bb{R}}

\def\bN {\bb{N}}

\def\ma{\mathrm{a}}
\def\mb{\mathrm{b}}

\def\gX{\Gamma\mkern-4.0mu X}
\def\gom{\Gamma\mkern-4.0mu \omega}
\def\gY{\Gamma\mkern-4.0mu Y}
\def\gF{\Gamma\mkern-4.0mu F}
\def\gQ{\Gamma\mkern-4.0mu Q}
\def\hn {D}

\def\sX{\slashed {X}}
\def\sgX{\slashed {\gX}}
\def\sY{\slashed {Y}}
\def\sgY{\slashed {\gY}}
\def\sh{\slashed {h}}
\def\sF{\slashed {F}}
\def\sgF{\slashed {\gF}}
\def\sQ{\slashed {Q}}
\def\sgQ{\slashed {\gQ}}
\def\sp{\slashed{\partial}}

\newcommand{\bea}{\begin{eqnarray}}
\newcommand{\eea}{\end{eqnarray}}

\begin{document}
\begin{titlepage}
\begin{center}
\vspace*{-1.0cm}
\hfill DMUS--MP--14/09 \\

\vspace{2.0cm} {\Large \bf 
Supersymmetry of AdS and flat backgrounds in M-theory } \\[.2cm]

\vskip 2cm
J. B.  Gutowski$^1$ and G. Papadopoulos$^2$
\\
\vskip .6cm

\begin{small}
$^1$\textit{Department of Mathematics,
University of Surrey \\
Guildford, GU2 7XH, UK. \\
Email: j.gutowski@surrey.ac.uk}
\end{small}\\*[.6cm]

\begin{small}
$^2$\textit{  Department of Mathematics, King's College London
\\
Strand, London WC2R 2LS, UK.\\
E-mail: george.papadopoulos@kcl.ac.uk}
\end{small}\\*[.6cm]

\end{center}

\vskip 3.5 cm
\begin{abstract}

We give a systematic description of all warped $AdS_n$ and $\bR^{n-1,1}$ backgrounds of M-theory and identify the a priori number of
supersymmetries that these backgrounds  preserve.
In particular, we show that $AdS_n$ backgrounds preserve $N= 2^{[{n\over2}]} k$ for $n\leq4$ and $N= 2^{[{n\over2}]+1} k$ for $4<n\leq 7$ supersymmetries while
 $\bR^{n-1,1}$ backgrounds preserve   $N= 2^{[{n\over2}]} k$ for $n\leq4$ and $N= 2^{[{n+1\over2}]} k$ for $4<n\leq7$,   supersymmetries.
Furthermore for $AdS_n$ backgrounds that satisfy the requirements for the maximum principle to hold, we show that the Killing spinors can be identified
with the zero modes of  Dirac-like operators on  $M^{11-n}$ coupled to fluxes thus establishing a new class of Lichnerowicz type theorems.
We also demonstrate that the Killing spinors of generic warped $AdS_n$ backgrounds do not factorize into  products of Killing spinors
on $AdS_n$ and  Killing spinors on the transverse space.

\end{abstract}

\end{titlepage}


\section{Introduction}

The $AdS$ backgrounds in eleven dimensions, following
the original work of Freund and Rubin \cite{FR}, have found widespread applications either as  supergravity compactifications, see \cite{duff, grana} for reviews,  or  more recently in the AdS/CFT correspondence as it applies to M-theory \cite{maldacena}. Because of this, there is an extensive literature in the
construction of $AdS$ backgrounds with emphasis on those preserving some of the supersymmetry of the underlying supergravity theory, for some selected references that include applications see
\cite{romans}-\cite{passias}.
Most of the computations made so far apply to special cases. These involve either
restrictions on the number of active fields  or a priori assumptions on the form of Killing spinors.

The purpose of this paper is to give a comprehensive description of all warped $AdS$ and flat   backgrounds of 11-dimensional supergravity,  $AdS_n\times_w M^{11-n}$ and $\bR^{n-1,1}\times_w M^{11-n}$, respectively,  making use of at most two assumptions\footnote{It is also assumed that the fields are sufficiently differentiable.}. One assumption
is that the backgrounds admit at least one supersymmetry. This applies to all cases that will be described in this paper. The second assumption, which applies to only  parts of the analysis,
is that  the fields and $M^{11-n}$ satisfy the requirements  necessary for  the maximum principle to apply. Such requirements include the assumptions
that the fields are smooth
 and that $M^{11-n}$ is compact without boundary.

 The focus of this paper is to count the number supersymmetries preserved by $AdS_n$ and $\bR^{n-1,1}$ backgrounds. In another paper \cite{madsgeom}, we
 shall describe the restrictions on the geometry of $M^{11-n}$  imposed by the KSEs and field equations of 11-dimensional
 supergravity. We have also included a brief description of the geometry of some of the $AdS_n$ backgrounds.

Our results include the identification of all a priori fractions of supersymmetry preserved by the $AdS_n\times_w M^{11-n}$ and $\bR^{n-1,1}\times_w M^{11-n}$ backgrounds.
In particular assuming that such backgrounds exist, we count the number of Killing spinors $N$ that they preserve.
We find that for $AdS_n$ backgrounds
\bea
N= 2^{[{n\over2}]} k~,~~~n\leq4~;~~~~N= 2^{[{n\over2}]+1} k~,~~~4<n\leq 7
\label{susyc}
\eea
where $k\in \bN_{>0}$. To prove this result for $AdS_2$ backgrounds we have assumed that the fields and $M^9$ satisfy the requirements of the maximum principle.
However for the rest of the $AdS_n$ backgrounds, this assumption is not necessary and the counting of supersymmetries stated is valid more generally.
The number of supersymmetries $N$ is further restricted. For example, restrictions arise from the classification results of \cite{maxsusy, n31, n30}
on backgrounds with maximal and near maximal supersymmetries. In addition all backgrounds preserving more than 16 supersymmetries are homogeneous
\cite{homogen}.
The results have been tabulated in table 1.

The number of supersymmetries preserved by  $AdS_n$ backgrounds can also be determined by counting the zero modes of
appropriate Dirac-like operators ${\mathscr {D}}{}^{(\pm)}$ coupled to fluxes on the transverse spaces $M^{11-n}$. This is achieved by proving  new types of Lichnerowicz
theorems which give a 1-1 correspondence between the Killing spinors and the zero modes of ${\mathscr {D}}{}^{(\pm)}$. To prove these
theorems, we assume that the fields and $M^{11-n}$ satisfy all the conditions required for the maximum principle to hold. As a result, we show that the number
of supersymmetries of $AdS_n$ backgrounds can be rewritten as
\bea
N= \ell(n) N_-~,
\label{susyg}
\eea
where $\ell(n)=2^{[{n\over2}]}$, $2\leq n \leq4$ and $\ell(n)=4$, $4<n\leq 7$ and $N_-=\mathrm{dim}\,\mathrm{Ker}\, {\mathscr {D}}{}^{(-)}$.

The explicit expression we obtain for the Killing spinors of $AdS_n$ backgrounds allows us to investigate whether they can be factorized
into the form $\epsilon=\xi\otimes \psi$, where $\xi$ is a Killing spinor on $AdS_n$ and $\psi$ is a Killing spinor on $M^{11-n}$. Such a factorization
has been widely used in the literature to solve the KSEs for such backgrounds.  We find that for this
factorization to hold an additional condition must be imposed on the Killing spinors which does not arise from the solution
of the KSEs on the spacetime.  Because of this we conclude that such a factorization does not hold for generic backgrounds. We demonstrate with an explicit example that for the correct
counting of supersymmetries, one should not assume the above factorization.

Our $AdS_n$ results can be adapted to $\bR^{n-1,1}$ or equivalently flat backgrounds. The latter arise in the limit where the AdS radius goes to infinity.  This limit
can be taken smoothly in all our local computations. But some of the regularity properties enjoyed by the $AdS_n$ backgrounds do not extend
in the limit. One of the main consequences of this is that the new Lichnerowicz theorems we have established for $AdS_n$ do not hold
for flat backgrounds. As a result, one cannot prove that $\bR^{1,1}$ backgrounds always preserve an even number of supersymmetries. However for the
rest of the $\bR^{n-1,1}$ backgrounds, we shall show that
\bea
N=2^{[{n\over2}]} k~,~~~2<n\leq 4~;~~~N=2^{[{n+1\over 2}]} k~,~~~4<n<8~.
\label{susyf}
\eea
 $N$ is further restricted. These results have been tabulated in table 2.   Observe that the number of supersymmetries  preserved by the $AdS_n$ and $\bR^{n-1,1}$ backgrounds differs. This is because the counting of linearly independent Killing spinors is different in the two cases.

The method we use to solve the KSEs for each  of the $AdS_n\times_w M^{11-n}$ backgrounds is based on the observation of \cite{adshor} that all such backgrounds
can be described as near horizon geometries, and that the KSEs can be integrated for M-horizons \cite{smhor, mhor, imhor}. In particular, we use this  to integrate the gravitino KSE of D=11 supergravity   along all
 $AdS_n$  directions  and then identify the independent KSEs on $M^{11-n}$.
The integration of KSEs along $AdS_n$ first involves the integration along two lightcone directions which arise naturally in the description of $AdS_n$ backgrounds as
near horizon geometries. This is achieved after decomposing the Killing spinor
as $\epsilon=\epsilon_++\epsilon_-$ according to some lightcone projections $\Gamma_\pm \epsilon_\pm=0$. This integration allows
the Killing spinors to be written  as $\epsilon_\pm=\epsilon_\pm(r,u, \phi_\pm)$, where $\phi_\pm$ depend only on the coordinates of the co-dimension
two subspace ${\cal S}$ defined by  $r=u=0$.  A key computation described in \cite{mhor, imhor}, after using Bianchi identities and field equations,  reveals that there are two remaining
independent KSEs one for each $\phi_\pm$,  which are derived from the naive restriction of the KSE equation of D=11 supergravity on ${\cal S}$.
This is sufficient to establish the formulae (\ref{susyc}) and (\ref{susyg}) for $AdS_2$ backgrounds. This is because such backgrounds are special cases
of horizons and so these formulae already follow from the results of \cite{smhor, mhor, imhor}. For $AdS_n$, $n>2$, to derive (\ref{susyc}), it is necessary to integrate
along the rest of the $AdS_n$ directions. It turns out that this is always possible at the cost of introducing  additional
algebraic KSEs on $M^{11-n}$. The Killing spinors can be expressed as $\phi_+=\phi_+(x, \tau_+, \sigma_+)$ and
$\phi_-=\phi_-(x, \tau_-, \sigma_-)$, where $x$
denote collectively the remaining $AdS_n$ directions and $\tau_\pm$ and $\sigma_\pm$ depend only on the coordinates of $M^{11-n}$. Moreover,
the remaining independent KSEs on $\tau_\pm$ and $\sigma_\pm$ consist of those that one can find by the naive restriction of the KSEs of $\phi_\pm$ on $M^{11-n}$, together with the
introduction of the above mentioned new algebraic KSEs one for each $\tau_\pm$ and $\sigma_\pm$.  Therefore, the independent KSEs can be arranged in four different sets, and each set contains two equations.

The derivation of the formula (\ref{susyc}) for  the $AdS_n$  backgrounds, $n>2$, is based on the observation that there are interwining Clifford
algebra operators which allow for a given solution to one of the four sets of KSEs to produce solutions to the other three sets. After taking
into account all such Clifford algebra operations, the formula (\ref{susyc}) can be established.

The proof of (\ref{susyg}) for $AdS_n$, $n>2$, requires  the choice of appropriate Dirac-like operators ${\mathscr {D}}{}^{(\pm)}$ on $M^{11-n}$. These are linear combinations
of the Dirac operators derived from the gravitino KSEs, and the new algebraic KSEs which arise on $M^{11-n}$ after integrating along the $AdS_n$ directions.
Then assuming that ${\mathscr {D}}{}^{(+)}\chi_\pm=0$, where $\chi_+$ is either $\sigma_+$ or $\tau_+$, one schematically establishes
\bea
D^2 \parallel \chi_+\parallel^2+ n A^{-1} \partial^i A \partial_i \parallel \chi_+\parallel^2= 2 \langle \mathbb{D}^{(+)} \chi, \mathbb{D}^{(+)} \chi_+\rangle+
2{9n-18\over 11-n} \parallel {\cal C}^{(+)}\chi_+\parallel^2~,
\label{maxmaxmax}
\eea
where $\mathbb{D}^{(+)}$ is the gravitino KSE appropriately modified by the new algebraic KSE and ${\cal C}^{(+)}$ is the new algebraic KSE. A similar formula
holds for the $\sigma_-$ and $\tau_-$ spinors. Provided that the requirements for the maximum principle to hold are satisfied by the fields and $M^{11-n}$,
i.e. for smooth fields and $M^{11-n}$ compact without boundary, one concludes that the only solutions to the above equation is that $\parallel\chi_+\parallel$
is constant and $\chi_+$ satisfies the KSEs. This demonstrates a new type of Lichnerowicz theorems for ${\mathscr {D}}{}^{(\pm)}$.
These theorems establish an 1-1 correspondence between the zero modes of ${\mathscr {D}}{}^{(\pm)}$ and  Killing spinors.
Using these, (\ref{susyg}) follows from (\ref{susyc}) and a counting of the multiplicity of zero modes of ${\mathscr {D}}{}^{(\pm)}$.  This is done by identifying the
Clifford algebra operators which commute with ${\mathscr {D}}{}^{(\pm)}$.

For the proof of the number of supersymmetries preserved by flat backgrounds $\bR^{n-1,1}$ in (\ref{susyf}), it suffices to investigate the KSEs that arise for the
$AdS_n$ backgrounds in the limit of infinite AdS radius. All our local computations are smooth in this limit and so these computations carry through to
all flat $n>2$ cases.   Using this, one can establish the independent KSEs on $M^{11-n}$ for flat backgrounds from those  of  the $AdS_n$ solutions. However there is
a difference in that the KSEs for $\sigma_\pm$ and $\tau_\pm$ spinors become identical at the limit of infinite $AdS$ radius. As a result
the  $\sigma_\pm$ and $\tau_\pm$ are not linearly independent and so the counting of supersymmetries differs from that of $AdS_n$ backgrounds.
After taking into account the Clifford algebra operators which intertwine between and those that commute with the KSEs, one can establish (\ref{susyf}).
Furthermore for flat backgrounds, some of the regularity assumptions used to establish the  Lichnerowicz type theorems for the $AdS_n$ backgrounds  do not hold.
As a result, there is no analogue of the formula (\ref{susyg}) for flat backgrounds.
For the same reason, one cannot establish that the number of supersymmetries preserved by $\bR^{1,1}$ backgrounds is even as this, for $AdS_2$ backgrounds,
requires the use of  global arguments.

This paper has been organized as follows. In section 2, we describe all $AdS_n$ and $\bR^{n-1,1}$ backgrounds as near horizon geometries and summarize
some of the results of \cite{mhor} on 11-dimensional horizons which are essential for the investigation that follows. In section 3,
we solve the KSEs for the $AdS_2$ backgrounds along the $AdS_2$ directions. In section 4 we demonstrate two new  Lichnerowicz type theorems for $AdS_2$ backgrounds utilizing
the maximum principle and verify that such backgrounds preserve an even number of supersymmetries.  In sections 5, 7 and  9, we solve the KSEs
for $AdS_3$, $AdS_4$  and $AdS_5$ backgrounds verifying (\ref{susyc}), respectively. In sections 6, 8 and 10,  we prove  new   Lichnerowicz type theorems for $AdS_3$, $AdS_4$  and $AdS_5$ backgrounds verifying (\ref{susyg}), respectively.  In section 11, we investigate the factorization of the Killing spinors of AdS backgrounds. In section 12, we prove (\ref{susyg}) for $\bR^{n-1,1}$ backgrounds. We state our conclusions and
describe some preliminary results on the geometry of $AdS_n$ and $\bR^{n-1,1}$ backgrounds in section 14.

\newsection{AdS backgrounds  and near horizon geometries}

\subsection{ Warped AdS and flat backgrounds}

To systematize the investigation  of warped AdS and Minkowski backgrounds in 11-dimensional supergravity,  $AdS_n\times_w M^{11-n}$ and $\bR^{n-1,1}\times_w M^{11-n}$,  respectively,
it is convenient to express them as near horizon geometries. This has already been suggested in \cite{adshor}, where a preliminary
analysis has been carried out.  In particular, the most general form of the metric and 4-form flux which includes all these backgrounds can be written in
 Gaussian null coordinates \cite{isen, gnull} as
\begin{eqnarray}
ds^2 &=& 2 {\bf{e}}^+ {\bf{e}}^- + \delta_{ij} {\bf{e}}^i {\bf{e}}^j=2 du (dr + r h - {1 \over 2} r^2 \Delta du)+ ds^2({\cal S})~,
\cr
F &=& {\bf{e}}^+ \wedge {\bf{e}}^- \wedge Y
+ r {\bf{e}}^+ \wedge d_h Y + X~,
\label{mhm}
\end{eqnarray}
where we have introduced the frame
\begin{eqnarray}
{\bf{e}}^+ = du~,~~~{\bf{e}}^- = dr + r h - {1 \over 2} r^2 \Delta du~,~~~
{\bf{e}}^i = e^i{}_J dy^J~;~~~g_{IJ}=\delta_{ij} e^i{}_I e^j{}_J~,
\label{nhbasis}
\end{eqnarray}
and
\begin{eqnarray}
ds^2({\cal S})=\delta_{ij} {\bf{e}}^i{\bf{e}}^j~,
\end{eqnarray}
is the metric on the space ${\cal S}$ transverse to the lightcone directions   given by $r=u=0$.
The dependence on the coordinates $r,u$ is given explicitly and $d_hY=dY-h\wedge Y$. In addition,
$\Delta$,  $h$,  $Y$ and $X$ are a 0-form, 1-form, 2-form and a 4-form on ${\cal S}$ and depend only on  the coordinates $y$
of ${\cal S}$.
We choose the frame indices  $i=1, 2, 3, 4, 6, 7, 8, 9, \sharp$ and we follow the conventions of \cite{imhor}.
The form of the fields in (\ref{mhm}) is the same as that of near horizon geometries investigated in \cite{smhor, mhor, imhor}.

There are many advantages to describe the $AdS$ backgrounds  in terms of near horizon geometries as in (\ref{mhm}). One of them is that some of the results obtained for
near horizon geometries in \cite{smhor, mhor, imhor} can be directly applied here, eg  one can use  the integrability of the light-cone directions for horizons to reduce the problem to the identification of the geometry  of near horizon
sections ${\cal S}$. This is a problem in Riemannian geometry which is easier to solve.
 To describe each $AdS_n\times_w M$ case separately, for $n=2,\dots, 11$, some additional restrictions  have to be imposed on the near horizon fields  (\ref{mhm}). This is because
the  $AdS_n\times_w M^{11-n}$ backgrounds  are invariant under the $SO(n-1,2)$ isometry group  of $AdS_n$.
Imposing this symmetry, we have the following.

\subsubsection{$AdS_2\times _w M^9$}

The metric and fluxes are

\begin{eqnarray}
ds^2 &=& 2 du (dr + r h - {1 \over 2} r^2 \Delta du)+ ds^2(M^9)~,
\cr
F &=& {\bf{e}}^+ \wedge {\bf{e}}^- \wedge Y
 + X~,
\label{mhmads2}
\end{eqnarray}
with
\bea
h=-2 A^{-1} dA=\Delta^{-1} d\Delta ~,~~~d_hY=0~,
\eea
where $A$, $Y$ and $X$ are a 0-form, 2-form  and a 4-form on $M^9={\cal S}$, respectively. Observe that $dh=0$.

\subsubsection{ $AdS_3\times _w M^8$}

The fields are
\begin{eqnarray}
ds^2 &=& 2 du (dr + r h )+ A^2 dz^2 +ds^2(M^8)~,
\cr
F &=& {\bf{e}}^+ \wedge {\bf{e}}^- \wedge dz \wedge Q  + X~,
\label{mhmads3}
\end{eqnarray}
with
\bea
h = -{2 \over \ell}dz -2 A^{-1} dA,~~~\Delta=0~,~~~Y=dz\wedge Q~,~~~d_hY=0
\eea
where $A, Q, X$ are a 0-form, 1-form and a 4-form of $M^8$, respectively, and depend only on the coordinates of $M^8$. $\ell$ is the radius of $AdS$.

\subsubsection{ $AdS_4\times _w M^7$}

The fields are

\begin{eqnarray}
ds^2 &=& 2 du (dr + r h )+ A^2 (dz^2+ e^{2z/\ell} dx^2) +ds^2(M^7)~,
\cr
F &=& e^{z/\ell}\, A^2 S\, {\bf{e}}^+ \wedge {\bf{e}}^- \wedge dz \wedge dx    + X~,
\label{mhmads4}
\end{eqnarray}
with
\bea
h = -{2 \over \ell}dz -2 A^{-1} dA,~~~\Delta=0~,~~~Y=e^{z/\ell} A^2 S\, dz\wedge dx ~,~~~d_hY=0
\eea
where $A, S, X$ are a 0-form, 0-form and a 4-form of $M^7$, respectively, and depend only on the coordinates of $M^7$. $\ell$ is the radius of $AdS$.

\subsubsection{ $AdS_n\times _w M^{11-n}$, $n>4$}

The fields of the rest of the backgrounds are

\begin{eqnarray}
ds^2 &=& 2 du (dr + r h )+ A^2 (dz^2+ e^{2z/\ell} \sum_{a=1}^{n-3}(dx^a)^2) +ds^2(M^{11-n})~,
\cr
F &=&   X~,
\label{mhmadsn}
\end{eqnarray}
with
\bea
h = -{2 \over \ell}dz -2 A^{-1} dA,~~~\Delta=0~,~~~Y=0~,
\eea
where $A, X$ are a 0-form and a 4-form of $M^{11-n}$, respectively, and depend only on the coordinates of $M^{11-n}$. $\ell$ is the radius of $AdS$.

Observe  that in all the above backgrounds $AdS_n\times_w M^{11-n}$, we have that ${\cal S}= H^{n-2}\times_w  M^{11-n}$, i.e. ${\cal S}$ is the warped product of hyperbolic (n-2)-dimensional space $H^{n-2}$ with the transverse space $M^{11-n}$ of $AdS_n$.

\subsubsection{ $\bR^{n-1,1}\times _w M^{11-n}$}

Another advantage of the universal ansatz (\ref{mhm}) is that it includes the warped  $\bR^{n-1,1}\times_w M^{11-n}$ backgrounds. These arise in the limit of large AdS radius, $\ell\rightarrow \infty$.
This limit is smooth in all our calculations and so our AdS results can be adapted to $\bR^{n-1,1}\times_w M^{11-n}$ backgrounds. However, many properties of the $AdS_n\times_w M^{11-n}$
backgrounds do not hold in the limit of $\bR^{n-1,1}\times_w M^{11-n}$ backgrounds and some care must be taken when taking this limit.

\subsection{Bianchi identities and field equations}

The fields of $AdS_n\times_w M^{11-n}$ backgrounds are  restricted by the field equations and Bianchi identities of 11-dimensional supergravity  \cite{julia}. Before
we proceed to apply these to each $AdS_n\times_w M$ background separately, it is convenient to decompose them for the universal ansatz (\ref{mhm}) along the light-cone
directions and the rest. In preparation for the applications to  $AdS_n\times_w M^{11-n}$, we shall also impose that $d_h Y=0$ and $dh=0$ which are satisfied for all these backgrounds.
The end result of the decomposition of the field equations and later the KSEs along the light-cone and ${\cal S}$ directions is to reduce the problem on ${\cal S}$
as the light-cone directions are integrable.
In particular, we find the following.

The decomposition of the  Bianchi identity   of $F$, $dF=0$, for the universal ansatz (\ref{mhm}) yields
\begin{eqnarray}
\label{clos}
dX=0 \ ,
\end{eqnarray}
i.e. $X$ is a closed form on ${\cal S}$.

Similarly, the field equation
\begin{eqnarray}
d \star_{11}  F -{1 \over 2} F \wedge F=0~,
\end{eqnarray}
of the 3-form gauge potential yields

\begin{eqnarray}
{\tilde{\nabla}}^i X_{i \ell_1 \ell_2 \ell_3} +
= h^i X_{i \ell_1 \ell_2 \ell_3} -
{1 \over 48} \epsilon_{\ell_1 \ell_2 \ell_3}{}^{q_1 q_2 q_3 q_4 q_5 q_6}
Y_{q_1 q_2} X_{q_3 q_4 q_5 q_6} \ \
\label{geq1}
\end{eqnarray}
and
\begin{eqnarray}
\label{ydiv}
{\tilde{\nabla}}^j Y_{ji}-{1\over 1152} \epsilon_{i}{}^{q_1 q_2 q_3 q_4 q_5 q_6 q_7 q_8} X_{q_1 q_2 q_3 q_4} X_{q_5 q_6 q_7 q_8}=0~,
\label{geq2}
\end{eqnarray}
where ${\tilde{\nabla}}$ is the Levi-Civita connection of the metric $ds^2({\cal S})$ on  ${\cal{S}}$.

The Einstein equation
\begin{eqnarray}
R_{MN} = {1 \over 12} F^2_{M N}
-{1 \over 144} g_{MN} F^2 \ .
\end{eqnarray}
of 11-dimensional supergravity  decomposes into a number of components. In particular along ${\cal{S}}$, one finds
\begin{eqnarray}
\label{ein1}
{\tilde {R}}_{ij} + {\tilde{\nabla}}_{(i} h_{j)} -{1 \over 2} h_i h_j = -{1 \over 2} Y^2_{i j}
+{1 \over 12} X^2_{ij}
+ \delta_{ij} \bigg( {1 \over 12} Y^2
-{1 \over 144} X^2 \bigg)~,
\end{eqnarray}
where  ${\tilde{R}}_{ij}$ is the Ricci tensor of ${\cal{S}}$.
The $+-$ component of the Einstein equation gives
\begin{eqnarray}
\label{einpm}
{\tilde{\nabla}}^i h_i = 2 \Delta + h^2 -{1 \over 3} Y^2
-{1 \over 72}X^2~.
\end{eqnarray}
Similarly, the $++$ and $+i$ components of the Einstein equation can be expressed as
\begin{eqnarray}
\label{einpp}
{1 \over 2} {\tilde{\nabla}}^i {\tilde{\nabla}}_i \Delta -{3 \over 2} h^i {\tilde{\nabla}}_i \Delta -{1 \over 2} \Delta
{\tilde{\nabla}}^i h_i + \Delta h^2
= 0~,
\nonumber \\
\end{eqnarray}
and
\begin{eqnarray}
\label{einpi}
 - {\tilde{\nabla}}_i \Delta + \Delta h_i
= 0\ ,
\nonumber \\
\end{eqnarray}
respectively. The last equation implies that if $\Delta\not=0$, which is the case for only the $AdS_2\times_w M^9$ backgrounds,  then $h=\Delta^{-1} d\Delta$.

As  has been explained in \cite{mhor},  the $++$ and the $+i$ components of the
Einstein equations,   ({\ref{einpp}}) and ({\ref{einpi}}), respectively, are not independent but they
 hold as a consequence of
({\ref{clos}}), the 3-form field equations ({\ref{geq1}}) and ({\ref{geq2}}) and
the components of the Einstein equation in ({\ref{ein1}}) and  ({\ref{einpm}}).
This does not make use of supersymmetry, or any other assumptions on
${\cal{S}}$.
Hence, the conditions on $ds^2({\cal S})$,  $\Delta$, $h$, $Y$ and $X$ simplify to
({\ref{clos}}), ({\ref{geq1}}), ({\ref{geq2}}), ({\ref{ein1}}) and ({\ref{einpm}}).

\subsection{Killing spinor equations}

Another advantage of using the universal ansatz (\ref{mhm}) is that one can apply the results obtained for near horizon geometries in \cite{mhor, imhor} to
integrate the KSEs
\begin{eqnarray}
\nabla_M \epsilon
+\bigg(-{1 \over 288} \sgF_M
+{1 \over 36} \sF_M\bigg) \epsilon =0~,
\nonumber \\
\label{d11kse}
\end{eqnarray}
of  11-dimensional supergravity \cite{julia}, where $\nabla$ is the spacetime Levi-Civita connection. In particular, it is known that the KSEs evaluated on the universal ansatz (\ref{mhm}) are integrable along the light-cone directions.
The analysis that follows is carried out in generality. In particular, we do not put any restrictions on the form of either the Killing spinor, or of the fields. Moreover,  we do not impose
any other additional restrictions,  like for example the bi-linear matching condition. Furthermore, the calculation is local and applies to all $AdS_n\times_w M$ backgrounds.

\subsubsection{Light-cone integrability and independent KSEs}

After integrating  along the light-cone directions following \cite{smhor, mhor}, one can write the Killing spinor $\epsilon$ as
\begin{eqnarray}
\epsilon = \epsilon_+ + \epsilon_-~,~~~\Gamma_\pm \epsilon_\pm =0~,
\label{dec}
\end{eqnarray}
with
\begin{eqnarray}
\label{ksp1}
\epsilon_+ = \phi_+ + u \Gamma_+ \Theta_- \phi_- , \qquad \epsilon_- = \phi_- + r \Gamma_-
\Theta_+\epsilon_+~,
\end{eqnarray}
where
\begin{eqnarray}
\Theta_\pm=\bigg({1 \over 4} \sh +{1 \over 288} \sX \pm {1 \over 12} \sY \bigg)~,
\end{eqnarray}
$\Gamma_\pm$ are light-cone projections, and $\phi_\pm = \phi_\pm (y)$ do not depend on $r$ or $u$ and depend only on the coordinates of ${\cal S}$.
For future reference, we write
\bea
\epsilon(\phi_-, \phi_+)=\phi_+ + u \Gamma_+ \Theta_- \phi_-+\phi_- + r \Gamma_-
\Theta_+\big(\phi_+ + u \Gamma_+ \Theta_- \phi_- \big)~.
\eea

Furthermore, one can show that the only remaining  independent KSEs are
\begin{eqnarray}
\label{ind1}
\nabla_i^{(\pm)}\phi_\pm\equiv {\tilde{\nabla}}_i \phi_\pm + \Psi^{(\pm)}_i \phi_\pm =0~,
\end{eqnarray}
where
\begin{eqnarray}
\Psi^{(\pm)}_i &=& \mp{1 \over 4} h_i -{1 \over 288} \sgX_i +{1 \over 36} \sX_i
\pm{1 \over 24} \sgY_i \mp{1 \over 6} \sY_i~,
\end{eqnarray}
and that if $\phi_-$ is a solution of the KSEs, $\nabla_i^{(-)}\phi_-=0$, then
\begin{eqnarray}
\label{etdef}
\phi_+ \equiv \Gamma_+ \Theta_- \phi_-
\end{eqnarray}
is also a solution, $\nabla_i^{(+)}\phi_+=0$. These results follow after substituting the spinor (\ref{dec}) into the KSEs (\ref{d11kse})
and after an extensive use of the Bianchi identities and field equations.

The substitution of  (\ref{dec}) into the KSEs (\ref{d11kse}) yields a large number of integrability conditions that are
satisfied after an extensive use of the Bianchi identities,  field equations and the independent KSEs (\ref{ind1}). However in what
follows, it is instructive to include two of these integrability conditions.  Although these are not independent, nevertheless they are useful in the solution
of the KSEs along the remaining AdS directions. After imposing $dh=d_hY=0$, which are satisfied for all $AdS_n\times_w M^{11-n}$ backgrounds, we have

\begin{eqnarray}
\label{cc1}
\bigg( {1 \over 2} \Delta
+2 \big({1 \over 4} \sh -{1 \over 288} \sX+{1 \over 12} \sY \big)
\Theta_+ \bigg) \phi_+ =0~,
\end{eqnarray}
and
\begin{eqnarray}
\label{cc3}
\bigg( -{1 \over 2} \Delta
+ 2 \big(-{1 \over 4} \sh +{1 \over 288} \sX +{1 \over 12} \sY\big)
\Theta_- \bigg)
\phi_- =0~.
 \end{eqnarray}

To summarize, the independent KSEs that have to be solved to find the supersymmetric $AdS_n\times_w M^{11-n}$ backgrounds are
(\ref{ind1}). In what follows, we shall demonstrate that these can be integrated along the remaining AdS directions in $AdS_n\times_w M^{11-n}$.

\newsection{$AdS_2$: Local analysis}

\subsection{Field equations}

In this case $M^9={\cal S}$,  the fields on ${\cal S}$ are
\bea
ds^2({\cal S})=ds^2(M^9)~,~~~\tilde F^{2}=Y~,~~~\tilde F^{4}=X~,~~~h=-2 A^{-1} dA \ .
\eea

Next, it follows from (\ref{einpi}) that  $h=\Delta^{-1} d\Delta$ and so
\bea
\Delta=\ell^{-2}A^{-2}~,
\eea
where $\ell$ is the radius of $AdS_2$.
Furthermore, the Bianchi identity for $F$ and $d_h Y=0$ imply that
\bea
dX=0~,~~~d(A^2 Y)=0~.
\eea
The remaining independent field equations are
\begin{eqnarray}
{\tilde{\nabla}}^i(A^2 X_{i \ell_1 \ell_2 \ell_3})
=  -
{A^2 \over 48} \epsilon_{\ell_1 \ell_2 \ell_3}{}^{q_1 q_2 q_3 q_4 q_5 q_6}
Y_{q_1 q_2} X_{q_3 q_4 q_5 q_6}~, \
\label{geq1ads2}
\end{eqnarray}
\begin{eqnarray}
\label{ein1ads2}
{\tilde {R}}_{ij} - {\tilde{\nabla}}_{i} \partial_{j} \log A^2 &-&{1\over2} \partial_i\log A^2 \partial_j\log A^2 = -{1 \over 2} Y^2_{ij}
+{1 \over 12} X^2_{ij}
\nonumber \\
&+& \delta_{ij} \bigg( {1 \over 12} Y^2
-{1 \over 144} X^2 \bigg)~,
\end{eqnarray}

\begin{eqnarray}
\label{einpmads2}
{\tilde{\nabla}}^i \partial_i\log A = -{A^{-2}\over \ell^2}  -2 \partial^i\log A \partial_i\log A  +{1 \over 6} Y^2
+{1 \over 144} X^2~,
\end{eqnarray}
and (\ref{geq2}) which remains unmodified.

\subsubsection{The warp factor $A$ is nowhere  vanishing}

Suppose that $M^9$ is path connected, the fields are smooth and  $A$ is not identical to zero but it may vanish at some points in $M^9$.  First let  $p\in M^9$ such that $A(p)=0$ and a point $q\in M^9$ such that $A(q)\not=0$. Without loss of generality we take $A(q)>0$, otherwise one can consider $-A$ for the argument that follows. Moreover consider a smooth path $\gamma$ with $\gamma(0)=q$ and $\gamma(1)=p$, and the function $f=A\circ \gamma$. As $A$ is continuous, there is a neighborhood $U$ of $q$ such that $A\vert_U\not=0$. Next consider, the set
\bea
V=\{a\in [0, \infty) \ : \ f(t)\not=0,~~~ \forall t\in [0, a]\}~.
\eea
$V$ is not empty as $0\in V$, and $W\subset V$ where $W=\{t\in [0,1] \ :  ~\gamma(t)\in U\}$. As $V$ is bounded by $1$, it has a  supremum $b$. Continuity requires that $f(b)=0$ and there is a sequence $\{t_n\}$ in $V$ which converges at $b$.

Next restricting (\ref{einpmads2}) at $\gamma$ and multiplying it with $f^2(t_n)\not=0$ for some finite $n$,  we find
\bea
f(t_n) \nabla^2 A+\partial_i A \partial^i A+{1\over \ell^2}+{f(t_n)^2\over 6} Y^2 +{f(t_n)^2\over 144} X^2=0~.
\label{vvv}
\eea
Now as $n\rightarrow \infty$, $f(t_n)$ approaches zero and becomes very small while the derivatives of $A$ and the fluxes are smooth and so their values are bounded when restricted on $[0,1]$ which  is compact. As a result, (\ref{einpmads2}) cannot be satisfied very close to $\gamma(b)$ as the term ${1\over \ell^2}$ which depends on the radius of AdS
cannot be arbitrary close to zero for any $\ell^2<\infty$. Thus $A$ cannot vanish.
  Later, we shall see that $A^2$ is related to the length of a parallel spinor in all $AdS_k$ cases
which again confirms that $A$ is nowhere vanishing.

\subsection{Killing spinor equations}

The KSEs on ${\cal S}=M^9$ are
\begin{eqnarray}
\label{ind1ads2}
\nabla_i^{(\pm)}\phi_\pm\equiv {\tilde{\nabla}}_i \phi_\pm + \Psi^{(\pm)}_i \phi_\pm =0~,
\end{eqnarray}
where
\begin{eqnarray}
\Psi^{(\pm)}_i &=& \pm{1 \over 4} \partial_i \log A^2-{1 \over 288} \sgX_i  +{1 \over 36} \sX_i
\pm{1 \over 24}  \sgY_i  \mp{1 \over 6} \sY_i~.
\end{eqnarray}
Furthermore from the general results on horizons,  if $\phi_-$ is a solution of the KSEs, $\nabla_i^{(-)}\phi_-=0$, then
\begin{eqnarray}
\label{etdefads2}
\phi_+ \equiv \Gamma_+ \Theta_- \phi_-~,
\end{eqnarray}
is also a solution, $\nabla_i^{(+)}\phi_+=0$, where
\begin{eqnarray}
\Theta_\pm=\bigg(-{1 \over 4} \sp\log A^2  +{1 \over 288} \sX \pm {1 \over 12} \sY \bigg)~.
\end{eqnarray}

It is not apparent that $\phi_+ = \Gamma_+ \Theta_- \phi_-\not=0$. In particular, $\phi_-$ can be in the kernel of $\Theta_-$. In the $AdS_2$ case, it has been shown in \cite{imhor}
that if $M^9$ is compact without boundary, then  $\mathrm{Ker}\, \Theta_-=\{0\}$ and so $\phi_+\not=0$ which leads to the enhancement of supersymmetry.

\newsection{$AdS_2$: Global analysis}

To prove that the number of supersymmetries preserved by $AdS_2$ backgrounds is even, it is required to establish
 Lichnerowicz type theorems relating the Killing spinors $\phi_\pm$,   $\nabla^{(\pm)}\phi_\pm=0$, to the zero modes of the Dirac-like operators
\begin{eqnarray}
\label{dirac1ads2}
{\cal{D}}^{(\pm)}=\Gamma^i {\tilde{\nabla}}_i  + \Psi^{(\pm)}  \ ,
\end{eqnarray}
where
\begin{eqnarray}
\Psi^{(\pm)} = \Gamma^i \Psi^{(\pm)}_i = \pm{1 \over 4} \sp\log A^2  +{1 \over 96} \sX \pm{1 \over 8} \sY~.
\end{eqnarray}
 Such theorems have been demonstrated in the context of near horizon geometries in \cite{imhor}.  Since the $AdS_2$ backgrounds are a special case
 of near horizon geometries, the result follows. However, there is a difference.   For $AdS_2$ backgrounds both  Lichnerowicz type theorems can be shown
 using the maximum principle which has an advantage relative to a partial integration formula used in \cite{imhor}.  This is because one also  finds
  a restriction on the length of the spinors.

\subsection{A Lichnerowicz type of theorem  for ${\cal D}^{(+)}$}

The proof of this theorem is identical to that in \cite{imhor} and so we shall not give details.    It is clear that if $\tilde\nabla^{(+)}\phi_+=0$, then $\phi_+$ is a zero mode of the ${\cal D}^{(+)}$ operator. To prove the converse, let us assume that  ${\cal D}^{(+)}\phi_+=0$.
 Then after using the field equations and Bianchi identities, one obtains
\begin{eqnarray}
\label{max2ads2}
{\tilde{\nabla}}^i {\tilde{\nabla}}_i \parallel \phi_+\parallel^2 +\partial^i\log A^2\, {\tilde{\nabla}}_i \parallel \phi_+\parallel^2
= 2 \langle {\tilde{\nabla}}^{(+)}{}^i \phi_+ , {\tilde{\nabla}}^{(+)}_i \phi_+ \rangle~.
\end{eqnarray}
Assuming that the requirements for applying of the maximum principle hold\footnote{From now on whenever we apply the maximum principle we shall assume that
 all the requirements on the fields and the associated manifold hold.} on the function $\parallel \phi_+\parallel^2$, eg $M^9$ is compact and the fields are smooth, one finds that
the only solutions to the above equation are
\bea
{\cal D}^{(+)}\phi_+=0\Longleftrightarrow {\tilde{\nabla}}^{(+)}_i \phi_+=0~,
\eea
and
\bea
\parallel\phi_+\parallel = \mathrm{const}~.
\eea
 This establishes the theorem.

\subsection{A Lichnerowicz theorem for ${\cal D}^{(-)}$}

The proof of the Lichnerowicz type  theorem for the ${\cal D}^{(-)}$ operator is done
in a similar fashion as that for ${\cal{D}}^{(+)}$.
It is clear that if $\tilde\nabla^{(-)}\phi_-=0$, then $\phi_-$ is a zero mode of the ${\cal D}^{(-)}$ operator. To prove the converse, let us assume that  ${\cal D}^{(-)}\phi_-=0$.
 Then after using the field equations and Bianchi identities, one obtains
\begin{eqnarray}
\label{max2ads2min}
{\tilde{\nabla}}^i {\tilde{\nabla}}_i \parallel \phi_-\parallel^2
+h^i {\tilde{\nabla}}_i \parallel \phi_-\parallel^2
+({\tilde{\nabla}}^i h_i)\parallel \phi_-\parallel^2
= 2 \langle {\tilde{\nabla}}^{(-)}{}^i \phi_- , {\tilde{\nabla}}^{(-)}_i \phi_- \rangle~.
\end{eqnarray}
For the $AdS_2$ solutions, $h=\Delta^{-1} d \Delta$, and this expression can be rewritten as
\begin{eqnarray}
\label{max2ads2min2}
{\tilde{\nabla}}^i {\tilde{\nabla}}_i \bigg( \Delta \parallel \phi_-\parallel^2 \bigg)
-h^i {\tilde{\nabla}}_i \bigg( \Delta \parallel \phi_-\parallel^2 \bigg)
= 2 \Delta \langle {\tilde{\nabla}}^{(-)}{}^i \phi_- , {\tilde{\nabla}}^{(-)}_i \phi_- \rangle~.
\end{eqnarray}
Since $\Delta$ is nowhere  zero applying the maximum principle on the function $\Delta \parallel \phi_-\parallel^2$, one concludes  that
\bea
{\cal D}^{(-)}\phi_-=0\Longleftrightarrow {\tilde{\nabla}}^{(-)}_i \phi_-=0
\eea
and
\bea
\Delta \parallel\phi_-\parallel^2 = \mathrm{const}~.
\eea
This establishes the 1-1 correspondence between Killing spinors and the zero modes of the Dirac-like operator ${\cal D}^{(-)}$.

\subsection{Counting supersymmetries}

One of the applications of the global analysis above is to prove that $AdS_2\times_w M$ backgrounds preserve $2k$, $0<k<15$,  supersymmetries.
The proof that $AdS_2\times_w M^9$ backgrounds preserve an even number of supersymmetries is similar to that for near horizon geometries established in \cite{imhor}.
The main point of the proof is that the ${\cal D}^{(-)}$  operator has the same principal symbol as the standard Dirac operator on the 9-dimensional
manifold $M^9$. As a result they have the same index which in this case vanishes,  as the index of the Dirac operator on compact without boundary odd dimensional manifolds is zero.  Moreover, one can show that the dimension of the Kernel of ${\cal D}^{(+)}$ operator is the same as that of $({\cal D}^{(-)})^\dagger$, the adjoint
of ${\cal D}^{(-)}$.  Now the number of supersymmetries $N$ preserved by the $AdS_2\times_w M$ backgrounds is
\bea
N&=&\mathrm{dim}\, \mathrm {Ker} {\cal D}^{(-)}+\mathrm{dim}\, \mathrm {Ker} {\cal D}^{(+)}=\mathrm{dim}\, \mathrm {Ker} {\cal D}^{(-)}+\mathrm{dim}\, \mathrm {Ker}({\cal D}^{(-)})^\dagger
\cr
&=&
2 \mathrm{dim}\, \mathrm {Ker} {\cal D}^{(-)}=2k ~,
\eea
where we have used that $\mathrm{dim}\, \mathrm {Ker} {\cal D}^{(-)}=\mathrm{dim}\, \mathrm {Ker} ({\cal D}^{(-)})^\dagger$ because the index of ${\cal D}^{(-)}$ is zero \cite{atiyah1}.

The restriction $0<k<15$ in the range of $k$ arises because of the results of \cite{n30} and \cite{maxsusy} which rule out the existence
of $AdS_2\times_w M$ backgrounds preserving 30 and 32 supersymmetries, respectively.

\newsection{$AdS_3$  : Local Analysis}

\subsection{Bianchi identities and Field equations}

The metric on ${\cal S}=\bR\times_w M^8$ and the form field strengths are
\bea
&&ds^2({\cal S})= A^2 dz^2 + ds^2(M^8)~,~~~ds^2(M^8)=\delta_{ij} \bbe^i \bbe^j~,
\cr
&&Y=A^{-1} \bbe^z\wedge Q~,~~~\tilde F=X~.
\eea
Next the Bianchi identity for $F$ implies that
\bea
d(A^2 Q)=0~,~~~dX=0~.
\eea

Let $\hn$ be the Levi-Civita connection\footnote{From here on $\hn$ will denote the Levi-Civita connection on $M^{11-n}$ and the latin indices $i,j,k, \ell$ are frame
 indices for $M^{11-n}$.} on $M^8$ and  $d\mathrm{vol}({\cal S})=\bbe^z\wedge d\mathrm{vol}(M^8)$. Then,
we find that the field equations ({\ref{geq1}}) and
({\ref{geq2}}) of the 4-form $F$ reduce on $M^8$ as
\bea
\label{gred1}
\hn^k X_{k i_1 i_2 i_3}
= -3 A^{-1} \hn^k A X_{k i_1 i_2 i_3} +{1 \over 24} A^{-1} \epsilon_{i_1 i_2
i_3}{}^{j_1 j_2 j_3 j_4 j_5}
Q_{j_1} X_{j_2 j_3 j_4 j_5}~,
\eea
and
\bea
\label{gred2}
\hn^k(A^{-1} Q_k)+{1 \over 1152}
\epsilon^{i_1 i_2 i_3 i_4
i_5 i_6 i_7 i_8}
X_{i_1 i_2 i_3 i_4}
X_{i_5 i_6 i_7 i_8} =0~.
\eea
We also reduce the Einstein equation ({\ref{ein1}}).
From the $zz$ component one obtains
\bea
\label{einred1}
A^{-1} \hn^k \hn_k A +2 A^{-2} \hn^k A
\hn_k A +{2 \over \ell^2} A^{-2}
= {1 \over 3} A^{-2}Q^2 +{1 \over 144} X^2~,
\eea
and from the $i, j$ component, one has
\bea
\label{rtensred}
{{R}}^{(8)}_{i j} =3 A^{-1} \hn_i \hn_j A
-{1 \over 2} A^{-2} Q_i Q_j +{1 \over 12}
X^2_{ij}
+ \delta_{i j} \bigg({1 \over 6} A^{-2}
Q^2 -{1 \over 144} X^2\bigg)~,
\eea
where ${{R}}^{(8)}$ denotes the Ricci tensor of $M^8$.

\subsubsection{The warp factor  $A$ is nowhere vanishing }

Before proceeding with the analysis of the supersymmetry, we
shall first examine whether or not $A$ can vanish
somewhere on $M^8$. For this, we use a similar set up as for the proof of $A\not=0$ for the $AdS_2$ case. In particular
after assuming that there are points $p$ and $q$ with  $A(p)=0$ and $A(q)\not=0$, constructing a path $\gamma$
between $p=\gamma(1)$ and $q=\gamma(0)$, and arguing the existence of a sequence $t_n$ such that $\lim_{n\rightarrow \infty} f(t_n)=0$ where $f(t_n)= A(\gamma(t_n))\not=0$,  we have
from  ({\ref{rtensred}}) and ({\ref{einred1}})
 that
\bea
f^2(t_n) {{R}^{(8)}} = -6 \hn^k A \hn_k A
-{6 \over \ell^2} +{11 \over 6} Q^2
+{7 \over 144} f(t_n)^2 X^2~,
\eea
and
\bea
f(t_n) \hn^k \hn_k A +2 \hn^k A \hn_k A
+{2 \over \ell^2} = {1 \over 3} Q^2+{1 \over 144}  f(t_n)^2  X^2~.
\eea
Eliminating the $Q^2$ term from the second equation using the first, we find
\bea
11 f(t_n) \hn^i \hn_i A+ 16 \hn^i A \hn_i A+{16\over \ell^2}= f^2(t_n) R^{(8)}+{1\over 36} f^2(t_n)  X^2~.
\label{limit2}
\eea
Assuming regularity for all the data and that  $M^8$ is path connected, an argument similar to that of the $AdS_2$ case implies that the above equation
cannot be satisfied if $A$ vanishes.  Therefore $A\not=0$ everywhere on $M^8$.

\subsection{Killing spinor equations}

To  reduce the KSEs
from ${\cal S}$  to $M^8$, we first decompose  (\ref{ind1}) along  the $z$-direction to find
\bea
\label{zder}
\partial_z \phi_\pm = \Xi^{(\pm)} \phi_\pm~,
\eea
where
\bea
\Xi^{(\pm)} = -{1 \over 2}
\Gamma_z \sp A
\mp{1 \over 2 \ell} +{A \over 288} \Gamma_z
\sX
\pm {1 \over 6}\sQ~,
\label{xiads3}
\eea
 $\partial_z \equiv {\partial \over \partial z}$ and $\Gamma_z$ denotes the frame gamma matrix from here on.
The  KSEs ({\ref{ind1}}) along the remaining directions can be written as
\bea
\label{tder}
D_i^{(\pm)} \phi_\pm=0~,
\eea
where
\bea
\label{tder2}
D_i^{(\pm)}=\hn_i\pm {1 \over 2} A^{-1} \hn_i A-{1 \over 288}\sgX_i+{1 \over 36}\sX_i\mp{1 \over 12} A^{-1} \Gamma_z
\sgQ_i  \pm {1 \over 6} A^{-1} \Gamma_z Q_i~.
\eea

Before proceeding further,  it is useful to evaluate the
following two integrability conditions
\bea
\label{intg1}
\Gamma^j \bigg(\hn_j \hn_i - \hn_i \hn_j \bigg)
\phi_\pm ={1 \over 2} {{R}^{(8)}}_{i j} \Gamma^j
\phi_\pm
\eea
and
\bea
\label{intg2}
\bigg(\partial_z \hn_i - \hn_i \partial_z \bigg)
\phi_\pm =0
\eea
where for both ({\ref{intg1}}) and ({\ref{intg2}}),
the KSE ({\ref{zder}}) and ({\ref{tder}}) are used to
expand out the LHS in terms of the fluxes.
After some involved computation, making use of
the field equations, one finds that
({\ref{intg1}}) and ({\ref{intg2}}) are in fact equivalent. This implies that there are non mixed integrability conditions between the $z$-direction
and the rest. Therefore the independent KSEs to solve are (\ref{zder}) and (\ref{tder}).

Next, we consider the algebraic conditions ({\ref{cc1}})
and ({\ref{cc3}}). Note that these arise from the integrability conditions along the light-cone directions and they are not independent, i.e.   they are implied by the remaining KSEs. Nevertheless,
 ({\ref{cc1}}), ({\ref{cc3}}), and together with ({\ref{zder}})
imply that
\bea
(\partial_z)^2 \phi_\pm \pm {1 \over \ell} \partial_z
\phi_\pm =0~.
\eea
This can be solved to yield
\bea
\label{pdef}
\phi_\pm = \sigma_\pm + e^{\mp {z \over \ell}} \tau_\pm~,
\eea
where
\bea
\partial_z \sigma_\pm = \partial_z \tau_\pm =0~,
\eea
and
\bea
\label{algcon}
\Xi^{(\pm)} \sigma_\pm =0, \qquad \Xi^{(\pm)} \tau_\pm = \mp {1 \over \ell} \tau_\pm~.
\eea

To summarize, in order to find a solution to the  KSEs for $AdS_3\times_w M^8$ backgrounds, it is sufficient to find spinors,
$\tau_\pm$ and $\sigma_\pm$ which depend only on the coordinates of $M^8$ satisfying
\bea
\label{transkse}
D^{(\pm)}_i \sigma_\pm &=&0~, \qquad   D^{(\pm)}_i \tau_\pm =0~,
\cr
{\cal{A}}^{(\pm)} \sigma_\pm &=&0, \qquad {\cal{B}}^{(\pm)} \tau_\pm =0~,
\eea
where $D^{(\pm)}_i$ is defined in (\ref{tder2}) and
\bea
{\cal{A}}^{(\pm)} =\Xi^{(\pm)}~,~~~{\cal{B}}^{(\pm)} =\Xi^{(\pm)}\pm {1 \over \ell}~.
\eea
Having given a solution to (\ref{transkse}), one can substitute it  in (\ref{pdef})  to construct the spinors $\phi_\pm$. In turn, $\phi_\pm$  will solve
the  KSEs (\ref{ind1}). Therefore to find the Killing spinors of the $AdS_3\times_w M^8$ backgrounds, it suffices to find the solutions of
(\ref{transkse}).

 \subsection{Counting supersymmetries}

 Unlike for  $AdS_2$ backgrounds, the counting of supersymmetries for $AdS_3$ backgrounds can be done using local geometry, and so additional
 requirements, like the conditions for the maximum principle to apply, are not necessary.
 For this we note that there is a 1-1 correspondence between the $\sigma_-$ and $\sigma_+$ solutions to the KSEs. Indeed given a $\sigma_-$ solution
 of ({\ref{transkse}}), then
\bea
\sigma_+ = A^{-1} \Gamma_+ \Gamma_z \sigma_-~,
\label{spsm}
\eea
automatically satisfies the KSEs, and conversely, if $\sigma_+$ is a solution then so is
\bea
\sigma_-=A \Gamma_- \Gamma_z \sigma_+~.
\label{smsp}
\eea
There is also an identical 1-1 correspondence between $\tau_+$ and $\tau_-$ Killing spinors.
Furthermore note that unlike for $AdS_2$ backgrounds,  $\Theta_-$ has a non-trivial kernel,
i.e. $\Theta_-\tau_-=0$.

Using the relation between the $(\sigma_-, \tau_-)$ and $(\sigma_+, \tau_+)$ spinors described above, we conclude  that the number of supersymmetries of the $AdS_3$ backgrounds is
\bea
N&=&2 \big(\mathrm{dim}\, \mathrm{Ker}(D^{(-)}, {\cal A}^{(-)})+ \mathrm{dim}\, \mathrm{Ker}(D^{(-)}, {\cal B}^{(-)})\big)~,
\cr
&=&2\big(\mathrm{dim}\, \mathrm{Ker}(D^{(+)}, {\cal A}^{(+)})+ \mathrm{dim}\, \mathrm{Ker}(D^{(+)}, {\cal B}^{(+)})\big)~.
\label{nads3}
\eea
Therefore all such backgrounds preserve an even number of supersymmetries establishing (\ref{susyc}).




\newsection{$AdS_3$  : Global Analysis}

The number of supersymmetries preserved by $AdS_3$ backgrounds can also be counted from the zero modes of Dirac-like operators on $M^8$
as stated in (\ref{susyg}). For this a 1-1 correspondence must be established between Killing spinors and the zero modes of these Dirac-like operators.  The proof
of this correspondence leads to new Lichnerowicz type theorems associated with the KSEs ({\ref{transkse}}).

\subsection{A new Lichnerowicz theorem for $\tau_+$ and $\sigma_+$}

One of the main difficulties in establishing (\ref{susyg}) is to choose appropriate Dirac-like operators on $M^8$.
The analysis is similar for $\sigma_+$ and $\tau_+$ spinors. Because of this, it is  straightforward to describe both cases at once.
Let $\chi$ denote either the $\sigma_+$ or the $\tau_+$ spinors and consider the operator
\bea
\label{modkse2}
\mathbb{D}^{(+)}_i\equiv \hn^{(+)}_i +k \Gamma_i{{\cal C}}{}^{(+)} ~,
\eea
where
\bea
{{\cal C}}{}^{(+)}=\big(-{1 \over 2}
A^{-1} \sp A
+{1 \over 288} \sX
+{1 \over 6} A^{-1} \Gamma_z \sQ
+c A^{-1} \Gamma_z \big)~,
\eea
 $c$ is a constant, with $c=-{1 \over 2 \ell}$
for $\chi = \sigma_+$ and $c={1 \over 2 \ell}$
for $\chi = \tau_+$, and $k$ is another constant which will later be set to $k=-{1 \over 8}$.
Therefore ${{\cal C}}{}^{(+)}={\cal A}^{(+)}$ if the operator  acts on $\sigma_+$, and ${{\cal C}}{}^{(+)}={\cal B}^{(+)}$ if the operator acts on $\tau_+$.
Next observe that if $\chi$ is a Killing spinor, then $\mathbb{D}^{(+)}\chi=0$.

The associated Dirac equation to (\ref{modkse2}) is
\bea
\label{moddirac}
\mathscr{D}^{(+)} \equiv \Gamma^i \mathbb{D}^{(+)}_i\chi &=& \Gamma^i D_i
+\bigg( ({1 \over 2}-4k) A^{-1}
\sp A
+({1 \over 72}+{k \over 36})\sX
\cr
&+&({5 \over 12}+{4 \over 3}k)A^{-1} \Gamma_z
\sQ + 8ck A^{-1} \Gamma_z \bigg)~.
\eea
In what follows, we shall demonstrate that the $\sigma_+, \tau_+$ Killing spinors are in 1-1 correspondence with the zero modes of $\mathscr{D}^{(+)}$.

It is clear that if $\chi$ is a Killing spinor, then it is a zero mode of $\mathscr{D}^{(+)}$. It remains to prove the converse. For this
assume that $\chi$ is a zero mode of $\mathscr{D}^{(+)}$, i.e. $\mathscr{D}^{(+)}\chi=0$, and that the fields of the theory satisfy the
field equations and the Bianchi identities. Then we shall demonstrate that for $k=-{1 \over 8}$
\bea
\label{max}
\hn^i \hn_i \parallel \chi \parallel^2
+3 A^{-1} \hn^i A \hn_i \parallel \chi \parallel^2
= 2 \langle \mathbb{D}^{(+)}{}^i\chi, \mathbb{D}^{(+)}_i \chi \rangle
+ {9 \over 4} \parallel {{\cal C}}{}^{(+)} \chi \parallel^2~.
\eea
An application of the maximum principle on the function $\parallel \chi \parallel^2$ reveals  that the zero modes of $\mathscr{D}^{(+)}$ are Killing spinors. Note that
as we have demonstrated $A$ is nowhere zero, it follows that $A^{-1}$ is smooth if $A$ is smooth.

To show (\ref{max}), we compute
\bea
\hn^i \hn_i \chi &=& \Gamma^k \hn_k\bigg(\Gamma^j \hn_j \chi \bigg)
+{1 \over 4}{{R}^{(8)}} \chi
\nonumber \\
&=& \Gamma^k \hn_k
\bigg(\big( (4k-{1 \over 2}) A^{-1}
\sp A
-({1 \over 72}+{k \over 36}) \sX
\nonumber \\
&-&({5 \over 12}+{4 \over 3}k)A^{-1} \Gamma_z
\sQ - 8ck A^{-1} \Gamma_z \big) \chi \bigg)
+{1 \over 4}{{R}^{(8)}} \chi~.
\eea
One then obtains
\bea
\label{auxv1}
{\rm Re} \ \langle \chi, \hn^i \hn_i \chi \rangle &=& \langle \chi , \bigg( (-{1 \over 2} +4k)
\hn^k (A^{-1} \hn_k A)+{1 \over 4} {\hat{R}}
+({5 \over 12}+{4 \over 3}k)\hn^k (A^{-1} Q_k)
\Gamma_z \bigg) \chi \rangle
\nonumber \\
&+& {\rm Re} \ \langle \chi, \Gamma^{i}
\bigg((-{1 \over 2}+4k) A^{-1} \sp A
-({1 \over 72}+{k \over 36}) \sX
\nonumber \\
&-&({5 \over 12}+{4 \over 3}k)
A^{-1} \Gamma_z \sQ-8ck A^{-1} \Gamma_z
\bigg) \hn_i \chi \rangle~.
\eea
We also write
\bea
\label{auxv2}
\langle \hn^i \chi, \hn_i \chi \rangle
&=& \langle\mathbb{D}^{(+)}{}^i\chi, \mathbb{D}^{(+)}_i \chi \rangle
+{\cal{F}}_1
\nonumber \\
&+& {\rm Re} \ \langle \chi ,
\bigg( -A^{-1} \hn^i A +{1 \over 144}
\sgX^{i}
+{1 \over 18} \sX^i
\nonumber \\
&-&{1 \over 6} A^{-1} \Gamma_z
\sgQ^i -{1 \over 3} A^{-1} \Gamma_z Q^i
+k \big(A^{-1} \sp A
-{1 \over 144} \sX
\nonumber \\
&+&{1 \over 3} A^{-1} \Gamma_z \sQ-2cA^{-1} \Gamma_z \big) \Gamma^i \bigg)
\hn_i \chi \rangle~,
\eea
where
\bea
{\cal{F}}_1 &=& \langle \chi , \bigg(
-{11 \over 32} A^{-2} \hn^i A \hn_i A
+{1 \over 384} A^{-1} \hn_{k_1}
A\, \sgX^{k_1}
-{5 \over 32} A^{-2} \Gamma_z Q_i \hn^i A
\nonumber \\
&+&{5 \over 24576} X_{k_1 k_2
k_3 k_4} X_{k_5 k_6 k_7
k_8} \Gamma^{k_1 k_2 k_3
k_4 k_5 k_6 k_7 k_8}
+{1 \over 1024} X_{k_1 k_2 i j}
X_{k_3 k_4}{}^{ij}
\Gamma^{k_1 k_2 k_3 k_4}
\nonumber \\
&-&{17 \over 3072} X^2
-{1 \over 128} A^{-1} \Gamma_z Q^i \sX_i
-{1 \over 16} A^{-2} Q^2
\nonumber \\
&+&{1 \over 384} c A^{-1} \Gamma_z \sX+{1 \over 16} c A^{-2} \sQ
-{1 \over 32 \ell^2} A^{-2} \bigg) \chi \rangle
\eea
is a term which is purely algebraic in
$Q, X, dA$ and $A$.
We then evaluate
\bea
{1 \over 2} \hn^i \hn_i
\parallel \chi \parallel^2 =
{\rm Re} \ \langle \chi, \hn^i \hn_i \chi \rangle
+ \langle \hn^i \chi, \hn_i \chi \rangle~.
\eea
We first consider the sum of the last two lines of
({\ref{auxv1}}) and the last three lines of ({\ref{auxv2}}).
We require that this should be written in the form
$\langle \chi , {\cal{F}}_2 \Gamma^k \hn_k
\chi \rangle$, where ${\cal{F}}_2$ is purely algebraic in
$Q, X, dA$ and $A$. This imposes the condition $k=-{1 \over 8}$ as was mentioned previously,
with
\bea
{\cal{F}}_2 = -{1 \over 384} \sX +{7 \over 8} A^{-1}\sp A
-{1 \over 8} A^{-1} \Gamma_z \sQ -{3 \over 4} c A^{-1} \Gamma_z~.
\eea

If $k=-{1 \over 8}$, one can then eliminate
almost all of the conditions involving $\hn_i \chi$ by making use of the Dirac equation. However, there is one term
of the type $\langle \chi , \hn^k A \hn_k  \chi \rangle$, which cannot be entirely removed
by such an elimination. This remaining term is equal
to $-{3 \over 2} A^{-1} \hn^i A \hn_i \parallel \chi \parallel^2$, and this will be retained.

We therefore set $k=-{1 \over 8}$, and expand out the
terms algebraic in the fluxes, making use of
the bosonic field equations to eliminate
the terms of the form $\hn^i X_{i k_1 k_2 k_3}$, $\hn^i (A^{-1}Q_i)$,
${{R}}^{(8)}$ and $\hn^i \hn_i A$ in terms
of terms algebraic in the fluxes and $dA$. Then after some computation, we obtain (\ref{max}).

To summarize, we have shown that for
 $k=-{1 \over 8}$
\bea
\mathscr{D}^{(+)}\chi=0 \Longleftrightarrow \mathbb{D}^{(+)}_i\chi=0~, \qquad
{{\cal C}}{}^{(+)} \chi=0~,
\nonumber \\
\eea
and furthermore
\bea
\parallel \chi\parallel={\rm const,}
\eea
as a consequence of (\ref{max}).
In particular, if $\sigma_+$, $\tau_+$  satisfy the Dirac equation ({\ref{moddirac}}) with $k=-{1 \over 8}$,
and $c=-{1 \over 2 \ell}$, $c={1 \over 2 \ell}$, respectively, then $\sigma_+, \ \tau_+$ satisfy the KSE
({\ref{transkse}}), and furthermore
\bea
\parallel \sigma_+ \parallel = {\rm const.},
\qquad \parallel \tau_+ \parallel = {\rm const.}
\eea
This concludes the proof of a new Lichnerowicz type theorem for the connection (\ref{modkse2}).

\subsection{A new Lichnerowicz theorem for $\tau_-$ and $\sigma_-$}

There is also an 1-1 correspondence between the $\tau_-, \sigma_-$ Killing spinors and the zero modes of a Dirac-like operator on $M^8$.
The proof follows from the relationship between $\tau_-, \sigma_-$ and $\tau_+, \sigma_+$ spinors given in (\ref{spsm}) and (\ref{smsp})
and the corresponding relations for $\tau_\pm$. Alteratively, one can repeat the analysis we have done
for $\sigma_+$ and $\tau_+$ now for $\sigma_-$ and $\tau_-$.

For this, we again treat both cases simultaneously by defining the operator
\bea
\mathbb{D}^{(-)}_i  \equiv \hn^{(-)}_i \psi+ k \Gamma_i  {{\cal C}}{}^{(-)} ~,
\eea
where
\bea
{\cal{C}}{}^{(-)} = -{1 \over 2}
A^{-1} \sp A    +{1\over 288}
\sX
- {1 \over 6} A^{-1} \Gamma_z \sQ+ \tilde c A^{-1} \Gamma_z~.
\eea
The operator $\mathbb{D}^{(-)}$ acts on the spinors $\psi$ such that for $\psi=\sigma_-$ one has ${\tilde{c}}={1 \over 2 \ell}$, and for $\psi=\tau_-$ one has ${\tilde{c}}=-{1 \over 2 \ell}$.
Equivalently, ${\cal{C}}{}^{(-)}= {\cal{A}}{}^{(-)}$ for ${\tilde{c}}={1 \over 2 \ell}$ and ${\cal{C}}{}^{(-)}= {\cal{B}}{}^{(-)}$ for ${\tilde{c}}=-{1 \over 2 \ell}$.  $k$ will be set to
$k=-{1 \over 8}$. It is clear from this that if $\psi$ is a Killing spinor, then $\mathbb{D}^{(-)}_i \psi=0$.

Next consider the Dirac-like operator
\bea
\label{dirac2}
{\mathscr {D}}{}^{(-)}   \equiv \Gamma^i \mathbb{D}^{(-)}_i =\Gamma^i D_i + \bigg(
{1 \over 96} \sX -{1 \over 4}
A^{-1} \Gamma_z \sQ -A^{-1} {\tilde{c}}
\Gamma_z \bigg)~.
\eea
Observe that if $\psi$ is a Killing spinor, then $\psi$ is a zero mode of ${\mathscr {D}}{}^{(-)} $.  To prove the converse, assume that ${\mathscr {D}}{}^{(-)}\psi=0 $ and after performing a  calculation similar to the one presented
for  ${\mathscr {D}}{}^{(+)} $, one can establish the formula
\bea
\label{max2}
\hn^i \hn_i \bigg( A^{-2} \parallel \psi \parallel^2 \bigg)
+3 A^{-1} \hn^i A \hn_i
\bigg( A^{-2} \parallel \psi \parallel^2 \bigg)
= 2 A^{-2} \langle \mathbb{D}^{(-)}{}^i \psi , \mathbb{D}^{(-)}_i \psi \rangle
\nonumber \\
+ {9 \over 4} A^{-2} \parallel {\cal{C}}{}^{(-)} \psi \parallel^2~,
\eea
for $k=-{1 \over 8}$.
In fact, for the purely algebraic terms not obtained from the bosonic field equations
and not involving $dA$, the resulting expressions are
identical, modulo the replacements $Q \rightarrow -Q$, $c \rightarrow {\tilde{c}}$.

It is clear that on applying the maximum principle
to ({\ref{max2}}), it follows that
\bea
{\mathscr {D}}{}^{(-)}  \psi=0 \Longleftrightarrow \mathbb{D}^{(-)}_i \psi=0, \qquad
{\cal{C}}{}^{(-)}\psi =0~,
\nonumber \\
\eea
 and furthermore that
 \bea
  \parallel A^{-1} \psi\parallel={\rm const.}
 \eea
Therefore, if $\sigma_-$, $\tau_-$  satisfy the Dirac equation ({\ref{dirac2}})
for $\tilde c={1 \over 2 \ell}$ and $\tilde c=-{1 \over 2 \ell}$, respectively, then $\sigma_-, \ \tau_-$ satisfy the KSEs
({\ref{transkse}}), and furthermore
\bea
 \parallel  A^{-1} \sigma_- \parallel= {\rm const.},
\qquad  \parallel  A^{-1} \tau_- \parallel= {\rm const.}
\eea
This together with the similar result in the previous section establish the 1-1 correspondence between
Killing spinors and zero modes of the Dirac-like operators.

\subsection{Counting supersymmetries again}

It is now straightforward to establish (\ref{susyg}). Provided that the requirements for the validity of the
new Lichnerowicz type theorems hold, the number of supersymmetries preserved by the $AdS_3$ backgrounds (\ref{nads3})  can be rewritten as
\bea
N&=&2 \big(\mathrm{dim}\, \mathrm{Ker}({\mathscr {D}}_{1/2\ell}^{(-)})+ \mathrm{dim}\, \mathrm{Ker}( {\mathscr {D}}_{-1/2\ell}^{(-)})\big)
\cr
&=&2\big(\mathrm{dim}\, \mathrm{Ker}({\mathscr {D}}_{-1/2\ell}^{(+)})+ \mathrm{dim}\, \mathrm{Ker}({\mathscr {D}}_{1/2\ell}^{(+)})\big)~,
\label{nads3c}
\eea
where the subscripts denote the values of  $\tilde c$ and $ c$.

\newsection{$AdS_4$: Local analysis}

\subsection{Field equations}

The metric and fluxes induced on ${\cal S}=H^2\times_w M^7$ from   $AdS_4\times_w M^{7}$ are
\bea
d\tilde s^2&=& (\bbe^z)^2+ (\bbe^x)^2 + ds^2(M^{7})~,~~~ds^2(M^{7})=\delta_{ij} \bbe^i \bbe ^j~,
\cr
Y&=& S\, \bbe^z\wedge \bbe^x~,~~~\tilde F=X~,
\eea
where
\bea
\bbe^z=A dz~,~~~\bbe^x= A e^{z/\ell} dx~.
\eea

Substituting these into the Bianchi identities and field equations for $F$, we find that
\bea
dX=0~,~~~ d(A^4 S)=0~,
\eea
and
\bea
\hn^k X_{ki_1i_2i_3} + 4 \hn^k A
X_{ki_1i_2i_3}=-{1\over 24} S \epsilon_{i_1i_2i_3}{}^{k_1k_2k_3k_4} X_{k_1k_2k_3k_4}~,
\eea
where we have set $d\mathrm{vol}({\cal S})=\bbe^z\wedge \bbe^x\wedge d \mathrm{vol}(M^7)$.

Furthermore, the Einstein equations  reduce on $M^7$ as
\bea
\hn^k \hn_k \log A=-{3A^{-2}\over \ell^2}-4 \partial_k \log A\partial^k \log A+{1\over 3} S^2+ {1\over 144} X^2~,
\eea
and
\bea
&& R^{(7)}_{ij}- 4 \hn_i\partial_j \log A-4 \partial_i \log A\partial_j \log A= {1\over12} X^2_{ij}
\cr
&&
\qquad +\delta_{ij} ({1\over 6} S^2-{1\over 114}X^2)~,
\eea
where $R^{(7)}$ is the Ricci tensor on $M^7$.
This completes the reduction of the Bianchi identities and field equations on $M^7$.

\subsubsection{The warp factor $A$ is nowhere vanishing}

Using a similar argument based of the field equations as that in the $AdS_3$ case, one can again demonstrate that $A$ is nowhere vanishing on $M^7$.
Again $A^2$ is related via the KSEs to the length of a parallel spinor confirming that it cannot vanish anywhere.

\subsection{Integrability of KSEs along AdS}

To integrate the independent KSEs (\ref{ind1}) along the remaining AdS directions, we shall first integrate along $\bbe^z$ and then the remaining $\bbe^x$ direction.
The integration along the $\bbe^z$ direction proceeds as in the previous $AdS_3\times_w M^8$  backgrounds and after using (\ref{cc1}) and (\ref{cc3}), we find that
\bea
\phi_\pm=\eta_\pm +e^{\mp z/\ell} \chi_\pm~,
\eea
where
\bea
\Xi^{(\pm)} \eta_\pm=0~,~~~\Xi^\pm \chi_\pm=\mp {1\over \ell} \chi_\pm~,
\eea
and
\bea
\Xi^{(\pm)}=-{1\over2} \Gamma_z\sp  A\mp {1\over 2\ell}  +{1\over 288} \Gamma_z A \sX \pm {1\over 6} A S \Gamma_{x}~.
\eea
The spinors $\eta_\pm$ and $\chi_\pm$ depend on the $x$ coordinate of $AdS_4$ and the coordinates of $M^7$. Next we integrate along the $x$ coordinate
to find that
\bea
\phi_+=\sigma_+-{1\over \ell} x \Gamma_x \Gamma_z \tau_++ e^{-z/\ell} \tau_+~,~~~\phi_-=\sigma_-+ e^{z/\ell} (-{1\over \ell} x \Gamma_x \Gamma_z\sigma_-+\tau_-)~,
\eea
where $\sigma_\pm$ and $\tau_\pm$ depend only on the coordinates of $M^7$. There are no further conditions to consider.

To summarize after integration over all AdS directions, the independent KSEs are
\bea
{D}_i^{(\pm)} \sigma_\pm=0~,~~~{D}_i^{(\pm)}\tau_\pm=0~,
\label{kseads41}
\eea
and
\bea
{\cal A}^{(\pm)} \sigma_\pm=0~,~~~{\cal B}^{(\pm)} \tau_\pm=0~,
\label{kseads42}
\eea
where
\bea
{D}_i^{(\pm)}&\equiv& D_i\pm {1\over 2} \partial_i \log A-{1\over 288} \sgX_i
+{1\over 36} \sX_i\pm {1\over12} S \Gamma_{i zx}~,
\cr
{\cal A}^{(\pm)}&\equiv& \Xi^{(\pm)}~,~~~{\cal B}^{(\pm)}\equiv \Xi^{(\pm)}\pm {1\over\ell} ~.
\eea
Therefore, the KSEs reduce to a set of parallel transport and algebraic equations on the transverse spaces $M^{7}$.

\subsection{Counting supersymmetries}

To find the supersymmetries preserved by the $AdS_4\times_w M^{7}$ backgrounds, notice that if $\sigma_\pm$ is a solution of the KSEs, then
\bea
\tau_\pm=\Gamma_{zx}  \sigma_\pm
\eea
is also a solution, and vice versa as $\Gamma_{zx}$ is invertible.   Furthermore, observe that if $\sigma_-, \tau_-$ is a solution, so is
\bea
\sigma_+=A^{-1} \Gamma_+ \Gamma_z \sigma_-~,~~~\tau_+=A^{-1} \Gamma_+ \Gamma_z \tau_-~.~~~
\label{pmrelads4}
\eea
and similarly, if  $\sigma_+, \tau_+$ is a solution, so is
\bea
\sigma_-=A \Gamma_- \Gamma_z \sigma_+~,~~~\tau_-=A \Gamma_- \Gamma_z \tau_+~.~~~
\label{mprelads4}
\eea
From the above relations, one concludes that the $AdS_4\times_w M^{7}$ backgrounds preserve
\bea
N= 4 \,\,\mathrm{dim} \mathrm {Ker}(D^{(-)}, {\cal A}^{(-)})
\label{nads4}
\eea
supersymmetries  for $0<\mathrm{dim} \mathrm {Ker}(D^{(-)}, {\cal A}^{(-)})\leq8$. This proves (\ref{susyc}).  For $\mathrm{dim}\, \mathrm {Ker}(D^{(-)}, {\cal A}^{(-)})=8$, there is a unique solution which is
locally isometric to $AdS_4\times S^7$.

\newsection{$AdS_4$: Global analysis}

\subsection{A new Lichnerowicz theorem for $\tau_+$ and $\sigma_+$}

To prove (\ref{susyg}), one has to establish a 1-1 correspondence between Killing spinors and zero modes of a Dirac-like operator.
To identify the appropriate Dirac-like operator, as for the $AdS_3$ case,
we take a linear combination of the   (\ref{kseads41}) and (\ref{kseads42}) KSEs and consider the modified gravitino KSE operator
\bea
\label{modkseads4}
\mathbb{D}^{(+)}_i  \equiv D^{(+)}_i
+k \Gamma_i{\cal C}^{(+)}~,
\eea
where
\bea
\label{tderads4}
{\cal C}^{(+)} =-{1 \over 2}
A^{-1} \Gamma^k \hn_k A
+{1 \over 288} \sX  +{1\over 6} S \Gamma_{zx}
+c A^{-1} \Gamma_z~.
\eea
$\mathbb{D}^{(+)}$ is acting on the spinors $\chi$ such that for $\chi=\sigma_+$, one has  $c=-{1\over2\ell}$,  and for $c={1\over2\ell}$, one has $\chi=\tau_+$. In what follows, one determines $k$ to formulate a maximum principle
for the length square $\parallel \chi\parallel^2$ of the spinor $\chi$.

The associated Dirac equation to (\ref{modkseads4}) is
\bea
{\mathscr D}^{(+)} &\equiv& \Gamma^i \mathbb{D}^{(+)}_i=\Gamma^i D_i+ \bigg[{1-7 k\over 2}\, A^{-1} \sp A
+{5+ 7 k\over 288} \sX
\cr
&&+ {7\over 4} S \Gamma_{zx}+7 c k  A^{-1} \Gamma_z\bigg]~.
\eea

Assuming now that ${\mathscr D}^{(+)} \chi=0$ and using the Bianchi identities and field equations, one can prove following steps similar to those described for the $AdS_3$ case that
\bea
&&D^2\parallel \chi\parallel^2+ 4 A^{-1} \partial^i A \partial_i \parallel \chi\parallel^2=2  \langle \mathbb{D}^{(+)}_i \chi, \mathbb{D}^{(+)}{}^i \chi \rangle
+ {36\over 7} \parallel {\cal C}^{(+)}\chi \parallel^2~,
\eea
provided that $k=-2/7$.

Applying the maximum principle on $\parallel \chi\parallel^2$, one finds that
\bea
{\mathscr D}^{(+)} \chi =0\Longleftrightarrow \mathbb{D}^{(+)}_i \chi=0~,~~~{\cal C}^{(+)} \chi=0
\eea
and in addition
\bea
\parallel\chi \parallel=\mathrm{const}~.
\eea
This establishes a 1-1 correspondence between the $\sigma_+$ and $\tau_+$ Killing spinors and zero modes of ${\mathscr D}^{(+)}$.

\subsection{A new Lichnerowicz theorem for $\tau_-$ and $\sigma_-$}

A Lichnerowicz type  theorem can be proven for the $\tau_-$ and $\sigma_-$ Killing spinors following similar steps to those described in the previous
section. Alternatively, one can use the relation between $\tau_-, \sigma_-$ and $\tau_+, \sigma_+$ spinors described in (\ref{pmrelads4}) and (\ref{mprelads4}). In either case,
assuming that $\psi$ is a zero mode of a modified Dirac-like operator ${\mathscr D}^{(-)}$, one can establish the equality
\bea
D^2\parallel A^{-1}\psi\parallel^2+ 4 A^{-1} \partial^i A \partial_i \parallel A^{-1} \psi\parallel^2&=&2  A^{-2}\langle \mathbb{D}^{(-)}_i \psi, \mathbb{D}^{(-)}{}^i \psi \rangle
\cr
&& + {36\over 7} A^{-2} \parallel {\cal C}^{(-)}\psi \parallel^2~,
\eea
where $\mathbb{D}^{(-)}= \hn^{(-)}+ k\Gamma_i {\cal C}^{(-)}$  is the modified gravitino KSE operator,
\bea
{\cal C}^{(-)}=-{1\over2} A^{-1} \sp A+ \tilde c A^{-1}\Gamma_z+{1\over288} \sX-{1\over 6} S \Gamma_{zx}~,
\eea
  $k=-2/ 7$ and $\tilde c={1\over 2\ell}$ for $\psi=\sigma_-$ and $\tilde c=-{1\over 2\ell}$ for $\psi=\tau_-$.  Moreover ${\mathscr D}^{(-)}=\Gamma^i \mathbb{D}^{(-)}_i$.  An application
of the maximum principle  implies that
\bea
{\mathscr D}^{(-)} \psi =0\Longleftrightarrow \mathbb{D}^{(-)}_i \psi=0~,~~~{\cal C}^{(-)} \psi=0~,
\eea
 and that
\bea
\parallel A^{-1} \psi\parallel=\mathrm{const}~.
\eea
The warp factor is therefore related to the length of the Killing spinors. Combining the results of this with the previous section,
we have established a 1-1 correspondence between the Killing spinors and the zero modes of Dirac-like operators on $M^7$.

\subsection{Counting supersymmetries}

Now we can establish (\ref{susyg}). It is simply a consequence of the Lichnerowicz type theorems proved in the previous two sections.  In particular, the number of supersymmetries
preserved by $AdS_4\times_w M^7$ backgrounds (\ref{nads4}) can now be written as
\bea
N= 4 \,\,\mathrm{dim}\,\mathrm{Ker}\, {\mathscr D}_{1/2\ell}^{(-)}~.
\eea
This formula can also be expressed in terms of any other three choices of Dirac-like operators.

\newsection{$AdS_n$ $n>4$: Local analysis}

\subsection{Bianchi identities and Field equations}

The metric and fluxes induced on ${\cal S}=H^{n-2}\times_w M^{11-n}$  from $AdS_n\times_w M^{11-n}$, $n>4$, are
\bea
d\tilde s^2&=& (\bbe^z)^2+\delta_{\ma\mb} \bbe^\ma \bbe^\mb + ds^2(M^{11-n})~,~~~ds^2(M^{11-n})=\delta_{ij} \bbe^i \bbe ^j~,
\cr
\tilde F&=&X~,
\eea
where
\bea
\bbe^z=A dz~,~~~\bbe^\ma= A e^{z/\ell} dx^a~.
\eea
The Bianchi identities  and field equations of the fluxes can be written as
\bea
D^k X_{ki_1i_2i_3}=-n A^{-1} D^k A X_{ki_1i_2i_3}~,~~~dX=0~.
\eea
Similarly after a decomposition, the independent Einstein field equations are
\bea
\label{lapein}
D^k \partial_k \log A=-{n-1\over \ell^2} A^{-2}- n \partial^k \log A\, \partial_k \log A+{1\over 144}  X^2~,
\eea
and
\bea
 R^{(11-n)}_{ij}- n D_i \partial_j\log A-n \partial_i\log A \partial_j\log A={1\over12} X^2_{ij}
-{1\over144} \delta_{ij} X^2~.
\eea
This completes the decomposition of Bianchi and the field equations on $M^{11-n}$.

\subsubsection{The warp factor $A$ is nowhere vanishing}

For $AdS_n\times_w M^{11-n}$ backgrounds, the warp factor does not vanish at any point of $M^{11-n}$. This follows
from the field equations as in all the previous cases.

\subsection{Integrability of KSEs along AdS}

To integrate the independent KSEs (\ref{ind1}) along the remaining AdS directions, we shall first integrate along $\bbe^z$ and then the remaining $\bbe^\ma$ directions.
The integration along the $\bbe^z$ direction proceeds as in the previous examples and after using (\ref{cc1}) and (\ref{cc3}, we find that
\bea
\phi_\pm=\eta_\pm +e^{\mp z/\ell} \chi_\pm~,
\eea
where
\bea
\Xi^{(\pm)} \eta_\pm=0~,~~~\Xi^{(\pm)} \chi_\pm=\mp {1\over \ell}  \chi_\pm~,
\eea
and
\bea
\Xi^{(\pm)}=-{1\over2}\Gamma_z \sp  A\mp {1\over 2\ell}   +{1\over 288}\Gamma_z A \sX~.
\eea
The spinors $\eta_\pm$ and $\chi_\pm$ depend on the $x^a$ and the coordinates of $M$. Next we integrate along the $x^a$ coordinates
to find that
\bea
\phi_+=\sigma_+-{1\over \ell} x^a \Gamma_\ma \Gamma_z \tau_++ e^{-z/\ell} \tau_+~,~~~\phi_-=\sigma_-+ e^{z/\ell} (-{1\over \ell} x^a \Gamma_\ma \Gamma_z\sigma_-+\tau_-)~,
\eea
where $\sigma_\pm$ and $\tau_\pm$ depend only on the coordinates of $M^{11-n}$.

Furthermore after integration over all AdS directions, the independent KSEs are
\bea
D^{(\pm)}_i \sigma_\pm=0~,~~~D^{(\pm)}_i \tau_\pm=0~,
\label{kseadsk1}
\eea

and
\bea
{\cal A}^{(\pm)} \sigma_\pm=0~,~~~{\cal B}^{(\pm)} \tau_\pm=0~,
\label{kseadsk2}
\eea

where
\bea
D^{(\pm)}_i&=&D_i \pm {1\over 2} \partial_i \log A-{1\over 288} \sgX_i
+{1\over 36} \sX_i~,
\cr
{\cal A}^{(\pm)}&=&\Xi^{(\pm)}~,~~~{\cal B}^{(\pm)}=\Xi^{(\pm)}\pm {1\over \ell}  ~.
\eea
Therefore, the investigation of solutions to the KSEs reduces to a set of parallel transport and algebraic equations on the transverse spaces $M^{11-n}$.

\subsection{Counting supersymmetries}

To find the supersymmetries preserved by the $AdS_n\times_w M^{11-n}$ backgrounds, notice that if $\sigma_\pm$ is a solution of the KSEs, then
\bea
\tau_\pm(v)=v^\ma\Gamma_z \Gamma_\ma \sigma_\pm
\eea
is also a solution for any constant vector $v$, and vice versa.  In addition if two vectors $v$ are orthogonal, then the associated spinors $\tau_\pm(v)$ are also
orthogonal and so linearly independent. Furthermore, observe that if $\sigma_-, \tau_-$ is a solution, so is
\bea
\sigma_+=A^{-1} \Gamma_+ \Gamma_z \sigma_-~,~~~\tau_+=A^{-1} \Gamma_+ \Gamma_z \tau_-~.~~~
\label{pmrel}
\eea
Similarly, if  $\sigma_+, \tau_+$ is a solution, so is
\bea
\sigma_-=A \Gamma_- \Gamma_z \sigma_+~,~~~\tau_-=A \Gamma_- \Gamma_z \tau_+~.~~~
\label{mprel}
\eea

As a result if $\sigma_-$ is Killing spinor, then
\bea
\Gamma_{\ma\mb} \sigma_-~,~~~\ma<\mb~,
\eea
are also Killing spinors of the same KSEs, i.e.  $D^{(-)}\Gamma_{\ma\mb} \sigma_-= \Xi^{(-)}\Gamma_{\ma\mb}\sigma_-=0$.  However not all $(\sigma_-, \Gamma_{\ma\mb} \sigma_-)$, $\ma<\mb$, are linearly independent.
Clearly if $\{\sigma_-, \Gamma_{\ma\mb} \sigma_-\}$, $\ma<\mb$ are mutually orthogonal, they are linearly independent. $\sigma_-$ is orthogonal to all $\Gamma_{\ma\mb} \sigma_-$, and $\Gamma_{\ma\mb} \sigma_-$ is orthogonal to $\Gamma_{\ma'\mb'} \sigma_-$, iff
the bilinear
\bea
\langle \sigma_-, \Gamma_{\ma\mb} \Gamma_{\ma'\mb'} \sigma_-\rangle=0~.
\label{ortho}
\eea
Using these, one can count the number of supersymmetries preserved by $AdS_n\times_w M^{11-n}$ backgrounds. A straightforward analysis reveals that if the solutions exist, $AdS_5\times_w M^6$
backgrounds preserve $8k$ supersymmetries and $AdS_6\times_w M^5$  backgrounds preserve $16k$ supersymmetries. Note that
\bea
\mathrm{dim}\, \mathrm{Ker}(D^{(-)}, \Xi^{(-)})= 2k~,~~~\mathrm{dim}\, \mathrm{Ker}(D^{(-)}, \Xi^{(-)})= 4k~,
\label{kindx}
\eea
respectively. In these two cases, the orthogonality condition (\ref{ortho}) is always automatically satisfied.

For  $AdS_7\times_w M^4$, (\ref{ortho})  gives a non-trivial restriction for $\ma, \mb, \ma'$ and $\mb'$ distinct.   To see this observe that $\Gamma_{\ma\mb\ma'\mb'}$ commutes with the
KSEs $(D^{(-)}, \Xi^{(-)})$ and so $\sigma_-$ can be taken to be in one of the two eigenspaces of $\Gamma_{\ma\mb\ma'\mb'}$ with eigenvalues $\pm1$. For $\sigma_-$ in one of the two  eigenspaces,
only 4 from the 7 spinors $\{\sigma_-, \Gamma_{\ma\mb} \sigma_-\}$, $\ma<\mb$, are linearly independent. As a result $AdS_7\times_w M^4$ can preserve $16k$ supersymmetries with $k$ given by  the second equation in
(\ref{kindx}).

\vskip2mm

To summarize, we have the following:

\begin{itemize}

\item[(i)] $AdS_5\times_w M^6$ backgrounds can preserve 8, 16 and 24 supersymmetries.  This follows from the results of  \cite{maxsusy} which rule out the existence of such backgrounds
preserving 32 supersymmetries.  It also turns out that there are no $AdS_5\times_w M^6$ preserving 24 supersymmetries.  The proof of this will be demonstrated in \cite{madsgeom}.

\item[(ii)] $AdS_6\times_w M^5$ backgrounds can preserve $16$ supersymmetries. The existence of such backgrounds preserving 32 supersymmetries is ruled out from the results
of \cite{maxsusy}.  It turns out that under some assumptions there are no backgrounds preserving 16 supersymmetries either, and a proof will be presented
in \cite{madsgeom}, see also \cite{passias}.

\item[(iii)] $AdS_7\times_w M^4$ can preserve $16$ and $32$ supersymmetries with the latter being  locally isometric
to the maximally supersymmetric $AdS_7\times S^4$ background of 11-dimensional supergravity.
  In fact all $AdS_7\times_w M^4$ solutions are locally isometric to $AdS_7\times S^4$.  This follows from the results of \cite{friedrich}  mentioned in \cite{hullfigueroa}.

\end{itemize}

Collecting all these results together, we establish (\ref{susyc})
for the rest of the $AdS_n$ backgrounds.


\newsection{$AdS_n$ $n>4$: Global analysis}

In this section, we prove (\ref{susyg}).  For this, we shall demonstrate new  Lichnerowicz type theorems associated with the KSEs (\ref{kseadsk1}) and (\ref{kseadsk2}).

\subsection{A new Lichnerowicz Theorem for $\tau_+$ and $\sigma_+$}

To show that there is a 1-1 correspondence between Killing spinors and zero modes of Dirac-like operators,
we take a linear combination of the   (\ref{kseadsk1}) and (\ref{kseadsk2}) KSEs and consider the modified gravitino KSE operator
\bea
\label{modkseadsk}
\mathbb{D}^{(+)}_i  \equiv D^{(+)}_i
+k \Gamma_i {\cal C}^{(+)} ~,
\eea
where
\bea
\label{tderadsk}
{\cal C}^{(+)}=-{1 \over 2}
A^{-1} \Gamma^k \hn_k A
+{1 \over 288} \sX
+c A^{-1} \Gamma_z~.
\eea
This operator acts on spinors $\chi$ such that for $\chi=\sigma_+$,  $c=-{1\over2\ell}$,  and for $\chi=\tau_+$,  $c={1\over2\ell}$.
The associated Dirac equation to (\ref{modkseadsk}) is
\bea
{\mathscr D}^{(+)}  &\equiv& \Gamma^i \hn_i+ \bigg[{1-k(11-n)\over 2}\, A^{-1} \sp A
+{n+1+ k(11-n)\over 288} \sX
\cr
&&+  c k (11-n) A^{-1} \Gamma_z\bigg]~.
\eea
Assuming now that ${\mathscr D}^{(+)} \chi=0$ and using the Bianchi identities and field equations, one can prove following steps similar to those described for the $AdS_3$ backgrounds that
\bea
D^2\parallel \chi\parallel^2+ n A^{-1} \partial^i A \partial_i \parallel \chi\parallel^2=2  \langle \mathbb{D}^{(+)}_i \chi, \mathbb{D}^{(+)}{}^i \chi \rangle
+2 {9n-18\over 11-n} \parallel {\cal C}^{(+)}\chi\parallel^2~,
\eea
provided that $k=2-n/ 11-n$.

For completeness, one has that
\bea
{\cal F}_1&=&\langle \chi, \bigg[-{11-3n+n^2\over4 (11-n)} (d\log A)^2+ {15-3n \over 288 (11-n)} \partial_{k}\log A\, \sgX^k
\cr
&&+ {63-18n \over 1152 (11-n)} X_{ij k_1k_2} X^{ij}{}_{k_3k_4}
\Gamma^{k_1k_2k_3k_4}
\cr
&&+ {-261+36n\over 3456 (11-n)} X^2
+{3n-6\over 144 (11-n)} c A^{-1} \Gamma_z \sX
\cr
&&- {(2-n)^2\over 4 (11-n)} A^{-2}\bigg]\chi\rangle
\eea
and
\bea
{\cal F}_2= {10n-n^2-7\over 2(11-n)} A^{-1} \sp A+{-15+3n\over288(11-n)}  \sX+ {(2-n) (9-n)\over 11-n} c A^{-1} \Gamma_z \ .
\eea

Applying the maximum principle one concludes that
\bea
{\mathscr D}^{(+)}\chi=0 \Longleftrightarrow \mathbb{D}^{(+)}_i\chi=0~,~~~{\mathcal C}^{(+)} \chi=0~,
\eea
  and that
\bea
\parallel \chi\parallel=\mathrm{const}~.
\eea
This proves the correspondence between the Killing spinors $\sigma_+$ and $\tau_+$ and the zero modes  of ${\mathscr D}^{(+)}$.

\subsection{A new Lichnerowicz Theorem for $\tau_-$ and $\sigma_-$}

The new Lichnerowicz theorem on the $\tau_-$ and $\sigma_-$ can be formulated either by  following a calculation similar to that of the
previous section or more simply by utilizing the relationship between the positive and negative lightcone chirality spinors and their associated KSEs
given in (\ref{pmrel}) and (\ref{mprel}).  The maximum principle formula can be written as
\bea
&&D^2\parallel A^{-1}\psi\parallel^2+ n A^{-1} \partial^i A \partial_i \parallel A^{-1} \psi\parallel^2=2  A^{-2}\langle \mathbb{D}^{(-)}_i \psi, \mathbb{D}^{(-)}{}^i \psi \rangle
\cr
&&
~~~~+2 {9n-18\over 11-n} A^{-2} \parallel {\cal C}^{(-)}\psi\parallel^2~,
\eea
where
\bea
\mathbb{D}^{(-)}_i= D^{(-)}_i+ k \Gamma_i {\cal C}^{(-)}~,
\eea
\bea
{\cal C}^{(-)}= -{1\over2} A^{-1} \sp A+ \tilde c A^{-1}\Gamma_z+{1\over288}  \sX~,
\eea
is the modified gravitino KSE,  $k=2-n/ 11-n$ and $\tilde c={1\over 2\ell}$ for $\psi=\sigma_-$ and $\tilde c=-{1\over 2\ell}$ for $\psi=\tau_-$.  The application
of the maximum principle  implies that
\bea
{\mathscr D}^{(-)}\psi=0 \Longleftrightarrow \mathbb{D}^{(-)}_i\psi=0~,~~~{\mathcal C}^{(-)} \psi=0~,
\eea
  and that
\bea
\parallel A^{-1} \psi\parallel=\mathrm{const}~.
\eea
Thus, the warp factor is proportional to the length of the Killing spinor.

\subsection{Counting supersymmetries}

It is now straightforward to prove (\ref{susyg}) for  $AdS_n\times_w M^{11-n}$, $n>4$
backgrounds. In particular this follows from (\ref{susyc}) and after setting
\bea
N_-=\mathrm{dim}\,\mathrm{Ker}\, {\mathscr D}_{1/2\ell}^{(-)}~.
\eea
This completes the proof of (\ref{susyg}) for all $AdS_n$ backgrounds.

\begin{table}
\centering
\fontencoding{OML}\fontfamily{cmm}\fontseries{m}\fontshape{it}\selectfont
\begin{tabular}{|c|c|}\hline
$AdS_n\times_w M^{11-n}$& $N$
 \\
\hline\hline

$n=2$&$2k, k<15$
\\
\hline
$n=3$&$2k, k<15$
\\
\hline
$n=4$&$4k, k\leq 8$
\\
\hline
$n=5$&$8, 16, 24$
\\
\hline
$n=6$&$16$
\\
\hline
$n=7$&$16, 32$
\\
\hline
\end{tabular}
\label{tab2}
\begin{caption}
{\small {\rm ~~The number of supersymmetries $N$ of $AdS_n\times_w M$ backgrounds are given. For $AdS_2\times_w M^{9}$, one can show that these backgrounds preserve an even number of supersymmetries
provided that they are smooth and $M^9$ is compact without boundary. For the rest, the counting of supersymmetries does not rely on the compactness of $M^{11-n}$.  The bounds in $k$ arise from the non-existence
 of supersymmetric solutions with near maximal supersymmetry. For the remaining fractions, it is not known whether there always exist backgrounds preserving the prescribed
 number of supersymmetries.
}}
\end{caption}
\end{table}

\section{Do the Killing spinors factorize?}

In many of the investigations of $AdS_n$ solutions of supergravity theories, it is assumed that the Killing spinors of the spacetime factorize
as
\bea
\epsilon=\xi\otimes \psi
\label{sansatz}
\eea
where $\xi$ is a Killing spinor on $AdS_n$ satisfying the Killing spinor equation
\bea
\nabla_\mu \xi+ \lambda \gamma_\mu \xi=0~,
\label{skse}
\eea
and $\psi$ is a spinor on the transverse space, and where $\mu$ are $AdS_n$ indices and $\gamma_\mu$ are $AdS_n$ gamma matrices. Then the ansatz (\ref{sansatz}) is substituted
 into the KSEs to derive the equations that must be satisfied by $\psi$. This factorization has been instrumental
in many of the AdS computations.  Here we shall examine whether the Killing spinors can  always be written in this way.

It is not straightforward to adapt our results to the ansatz (\ref{sansatz}) for the spinors stated above. This is because of the ambiguities that one encounters when decomposing the spacetime
gamma matrices in terms of those of $AdS$ and transverse space as well as the differences in the choice of coordinates on $AdS_n$ and spacetime frame. However we can restrict our Killing spinors on $AdS_n$
and examine whether they satisfy (\ref{skse}). Indeed, if the Killing spinors factorize as in (\ref{sansatz}) and $\psi$ is taken to be a constant, then $\epsilon$ solves (\ref{skse}).

Consider first the $AdS_2$ case. Observe that after using the integrability condition (\ref{cc3}) the Killing spinors can be written as
\bea
\epsilon= \phi_++\phi_-+ u \Gamma_+\Theta_-\phi_-+ r \Gamma_- \Theta_+ \phi_++ {\Delta\over2} r u \phi_-~,
\eea
where $\phi_\pm$ lie in the ${\bf 16}$ Majorana representation of $Spin(9)$, the spin group of $M^9$, and they are localized on $M^9$.   $\phi_\pm$ are the candidates to be identified with $\psi$.
On imposing the condition ({\ref{skse}}) we find
\bea
\label{ads2fact1}
\bigg(\Theta_\pm + \lambda \bigg) \phi_\pm =0, \quad \lambda^2={\Delta \over 4} \ .
\eea
However, these conditions are incompatible with the KSEs.
To see this, observe that the horizon Dirac equations on ${\cal{S}}$
can be written as
\bea
\label{ads2fact2}
\Gamma^i {\tilde{\nabla}}_i \psi_\pm +3 \Theta_\pm \psi_\pm =0
\eea
where $\psi_+= \Delta^{-1} \phi_+$, $\psi_- = \Delta^{-{1 \over 2}} \phi_-$. Then ({\ref{ads2fact2}}) implies that
\bea
\label{asd2fact3}
\int_{M^9} \langle \psi_\pm, \Theta_\pm \psi_\pm \rangle =0 \ .
\eea
However, substituting ({\ref{ads2fact1}}) into ({\ref{asd2fact3}})
then leads to a contradiction. Hence the $AdS_2$ background spinors
cannot factorize as in ({\ref{sansatz}}).

Next let us turn to the $AdS_3$ case. After applying the integrability condition (\ref{cc3}) and using the algebraic KSEs (\ref{algcon}), the Killing spinor can be written as
\bea
\epsilon=\sigma_++ \sigma_-+ e^{-{z\over\ell}} \tau_++ e^{z\over\ell} \tau_--{u\over\ell} A^{-1} \Gamma_{+z} \sigma_--{r\over\ell} A^{-1} e^{-{z\over\ell}} \Gamma_{-z} \tau_+~.
\eea
Observe that $\sigma_\pm$ and $\tau_\pm$ are in the ${\bf 16}$ Majorana (but not Weyl) representation of $Spin(8)$, the spin group of $M^8$, and they are localized
on $M^8$. Next let us restrict $\epsilon$ on $AdS_3$ by treating $\sigma_\pm$ and $\tau_\pm$ as integration constants, i.e. suppress  all the dependence of $\epsilon$ on
the coordinates of $M^8$, and investigate whether the resulting spinor solves the KSE (\ref{skse}) on $AdS_n$ for some choice of $\lambda$.  Substituting $\epsilon$ into (\ref{skse}), one finds that
it solves the KSE if and only if
\bea
{1\over2\ell} A^{-1} \Gamma_z \sigma_\pm \pm \lambda \sigma_\pm=0~, ~~~{1\over2\ell} A^{-1} \Gamma_z \tau_\pm\mp \lambda \tau_\pm=0~.
\label{solskse}
\eea

Let us focus on the condition (\ref{solskse}) for $\sigma_+$.
This condition can be solved by imposing the projection $\Gamma_z\sigma^\pm_+=\pm \sigma_+^\pm$ with $\sigma_+=\sigma_+^++ \sigma_+^-$
and $\lambda=\mp {1\over2\ell} A^{-1}$. Therefore $\epsilon$ solves (\ref{skse}) if and only if one of the components of $\sigma_+$, $\sigma_+^+$ or $\sigma_+^-$,
is set to zero. However, this is an additional condition on $\sigma_+$ which does not arise from the KSEs as applied to $AdS_3$ backgrounds.
In fact since $\Gamma_z$ does not commute with the KSEs, one cannot consistently set $\sigma_+^+$ or $\sigma_+^-$ to zero without imposing
an additional restriction on the fields.  A similar analysis can be done for the remaining three cases.
Therefore for generic $AdS_3$ backgrounds, the Killing spinors cannot be written as in (\ref{sansatz}) satisfying (\ref{skse}).

Next let us consider the same question for $AdS_n$ backgrounds for $n>3$.  The Killing spinors for these backgrounds, after using the integrability condition
(\ref{cc3}) and the algebraic KSEs ${\cal A}^{(\pm)} \sigma_\pm={\cal B}^{(\pm)}\tau_\pm=0$, can be written as
\bea
\epsilon&=&\sigma_++\sigma_-+ e^{-{z\over\ell}} \tau_+ + e^{z\over\ell} \tau_-
\cr
&&-{1\over\ell} \big( u  A^{-1} \Gamma_{+z} \sigma_-+ r  A^{-1} e^{-{z\over\ell}} \Gamma_{-z} \tau_+
+ \sum_\ma x^\ma \Gamma_\ma \Gamma_z (\tau_++ e^{z\over\ell} \sigma_-)\big)~.
\eea
Again  $\epsilon$ depends on four linearly independent spinors $\sigma_\pm$ and $\tau_\pm$. As in the $AdS_3$ case substituting\footnote{In the KSE (\ref{skse}) we use the spacetime gamma
 matrices. The same result holds if one uses the AdS gamma matrices after one decomposes $\sigma_\pm$ and $\tau_\pm$ in terms of AdS and transverse spinors.}
 $\epsilon$ into  (\ref{skse}),
one finds that $\sigma_\pm$ and $\tau_\pm$ must satisfy (\ref{solskse}).
 Repeating the argument we developed for the $AdS_3$ case, one concludes that $\epsilon$
does not factorize\footnote{This questions the generality of several results that have been obtained
in the literature using the factorization ansatz (\ref{sansatz}). How this additional
assumption affects each computation is not apparent and the claims about the generality of several results have to be re-examined.}  as in (\ref{sansatz}).

To illustrate the importance of keeping both components of $\sigma_+$, $\sigma_+^+$ and $\sigma_+^-$, under the projection with $\Gamma_z$
for the correct counting of supersymmetries, let us consider the $AdS_7\times S^4$ background. One solves the algebraic KSEs, ${\cal A}^{(+)}\sigma_+=0$,
after imposing the projection\footnote{The other projection $\Gamma_z\Gamma^{[4]}\sigma_+=-\sigma_+$ can be treated in a similar way.}
\bea
\Gamma_z\Gamma^{[4]}\sigma_+=\sigma_+~,
\eea
where $\Gamma^{[4]}$ is the Clifford element associated with the volume form of $S^4$. This projection reduces the number of components
of $\sigma_+$ from 16 to 8. Then the gravitino KSE can be solved without additional conditions. Using the intertwining Clifford algebra
operators to find solitions to the KSEs for $\sigma_-$ and $\tau_\pm$, one concludes that $AdS_7\times S^4$ preserves 32 superymmetries
as expected.  However if one in addition imposes $\Gamma_z \sigma^\pm_+=\pm \sigma_+^\pm$ in order for the Killing spinor $\epsilon$ to satisfy
(\ref{skse}), then one would incorrectly conclude that $AdS_7\times S^4$ preserves 16 supersymmetries.

We have demonstrated that the non-factorization of the Killing spinor as in (\ref{sansatz}) is crucial for the correct counting of supersymmetries of these backgrounds.
On the other hand, the effect of the non-factorization of $\epsilon$ on the geometry of the transverse spaces $M^{11-n}$ is more subtle. A preliminary
investigation \cite{madsgeom} suggests that in some cases it has an effect but in some other cases it does not. Therefore it is required to investigate each case separately.

\section{$\bR^{n-1,1}$  M-theory backgrounds}

As  has been mentioned in the introduction the warped flat backgrounds $\bR^{n-1,1}\times_w M^{11-n}$ can be investigated
in the context of $AdS_n$ backgrounds after taking the AdS radius $\ell$ to infinity.  All our local computations are smooth
in this limit.  As a result, one can solve the KSEs, decompose the field equations and Bianchi identities as in the $AdS_n$
case and then take the limit $\ell\rightarrow \infty$ to derive the corresponding formulae for flat backgrounds.
However in the limit, not all properties of $AdS_n$ backgrounds carry through. In particular the solutions to the KSEs will be re-examined as
the process of integration yields powers of $\ell$ which alters
the conclusions somewhat. To emphasize another difference, it is known for sometime
that there are no smooth warped compactifications of 11-dimensional supergravity \cite{maldacena2}.  This is unless one appeals to M-theory
and includes higher curvature corrections, due for example to anomaly cancellation, and/or additional brane charges.
All these backgrounds are constructed starting from those of 11-dimensional supergravity and then appropriately correcting them. Since the 11-dimensional
backgrounds are the starting point for such computations,   we shall focus on these here.

\subsection{The warp factor is not nowhere vanishing}

One of the key properties of $AdS_n$ backgrounds is that the warp factor $A$ is nowhere vanishing.  This is not the case for
flat backgrounds. For flat backgrounds with non-trivial fluxes and under certain conditions, $A$ must always vanish somewhere\footnote{From now one, we shall assume that $A$ is non-vanishing
on an open subset of $M^{11-n}$ and restrict our analysis to that subset.}. This follows from \cite{maldacena2}. To see this focus
on the $\bR^{1,1}$ backgrounds and in particular in the field equation
(\ref{einpmads2}). After taking the limit $\ell\rightarrow \infty$  and assuming  that the conditions on the fields and $M^9$ for the maximum principle to apply hold, one
finds that the only solution
to this equation is  $A$ constant and $Y=X=0$, i.e. all the fluxes vanish, $F=0$. Therefore for compact $M^9$ and smooth fields, $A$ must vanish somewhere on $M^9$.
A similar analysis can be done for the other flat backgrounds.

\subsection{Counting supersymmetries}

Because our global techniques do not straightforwardly apply to flat backgrounds, we shall focus on the counting of their supersymmetries based only
on the local solution of the KSEs.

\subsubsection{$\bR^{1,1}$  backgrounds}

It is instructive to re-examine the integration of the KSEs in this case, in the limit that $\ell\rightarrow \infty$, $\Delta=0$. Using the
expression for the Killing spinor in (\ref{ksp1}) and the integrability conditions (\ref{cc1}) and (\ref{cc3}), we find that
\bea
\epsilon_+=\phi_++ u \Gamma_+\Theta_-\phi_-~,~~~\epsilon_-=\phi_-+r \Gamma_-\Theta_+ \phi_+~.
\eea
Therefore $\epsilon$ can depend on the lightcone coordinates $r, u$.
Of course this dependence relies on $\phi_\pm \notin \mathrm{Ker} \Theta_\pm$.  However, it should be stressed that
$\phi_\pm$ satisfy automatically the integrability conditions (\ref{cc1}) and (\ref{cc3}) as a consequence of the remaining KSEs (\ref{ind1ads2})  on $M^9$, the field equations
and Bianchi identities.   Therefore, the dependence on the lightcone coordinates depends crucially on been able to find solutions of
(\ref{ind1ads2}) which satisfy (\ref{cc1}) and (\ref{cc3}) but do not lie in the kernel of $\Theta_\pm$.

The proof that we gave for $AdS_2$ backgrounds to preserve an even number of supersymmetries is not automatically valid for flat backgrounds.
 This is because it has been based on global considerations which are no longer valid for $\bR^{1,1}$ solutions. Nevertheless, it is expected that some $\bR^{1,1}$ backgrounds
exhibit supersymmetry enhancement. Indeed suppose that $\phi_-$ is a Killing spinor. Then we have shown by a local computation
that $\phi_+=\Gamma_+\Theta_-\phi_-$ is also a Killing spinor.  Thus if $\phi_-\notin \mathrm{Ker}\, \Theta_-$, then such backgrounds
will admit at least two supersymmetries.

\subsubsection{$\bR^{2,1}$  backgrounds}

To determine the Killing spinors for this backgrounds, let us reexamine the solutions of the KSEs (\ref{ksp1}) along the additional $z$ direction
of $\bR^{2,1}$.  The relevant equation is given in (\ref{zder}). To continue observe that the $\Xi^{(\pm)}$ given in
(\ref{xiads3}) satisfy
\bea
\big(\Xi^{(\pm)}\big)^2\phi_\pm=0~,
\label{nil3}
\eea
in the limit $\ell\rightarrow \infty$ as a consequence of the integrability conditions (\ref{cc1}) and (\ref{cc3}). Thus
the most general solution can be written as
\bea
\phi_\pm=\sigma_\pm+ z\Gamma_z A \Theta_{\pm}\tau_\pm~,~~~\Xi^{(\pm)}(\sigma_\pm-\tau_\pm)=0~,~
\eea
where we have used that $\Xi^{(\pm)}=A \Gamma_z\Theta_{\pm}$.  Using again (\ref{nil3}), the Killing spinors can be expressed as
\bea
\epsilon_+=\sigma_++ u \Gamma_+\Theta_-\sigma_-+z\Gamma_z A \Theta_{+}\sigma_+~,~~~\epsilon_-=\sigma_-+r \Gamma_-\Theta_+ \sigma_++z\Gamma_z A \Theta_{-}\sigma_-~.
\eea
Therefore the Killing spinors can depend on the coordinates\footnote{As we shall see this is the case for the rest of the $\bR^{n-1,1}$ backgrounds.
 It would be of interest to find  solutions that exhibit this property, which is allowed from consideration of the KSEs, in order to test whether  field equations and
 Bianchi identities  impose additional conditions which remove this dependence.} of $\bR^{2,1}$. But this depends crucially on $\sigma_\pm\notin \mathrm{Ker}\, \Xi^{(\pm)}$
even though they are in the kernel of $\big(\Xi^{(\pm)}\big)^2$ because of (\ref{nil3}).

Furthermore a direct observation of the KSEs on $M^8$ (\ref{transkse}) reveals that  if $\sigma_-$ is a solution, then
\bea
\sigma_+= A^{-1} \Gamma_+\Gamma_z \sigma_-
\eea
is also a solution, and vice versa if $\sigma_+$ is a solution, then so is
\bea
\sigma_-= A \Gamma_- \Gamma_z \sigma_+~.
\eea
Therefore all $\bR^{2,1}$ backgrounds preserve an even number of supersymmetries confirming (\ref{susyf}).

\subsubsection{$\bR^{3,1}$  backgrounds}

As in the previous case, it is straightforward to observe that the solution to the KSEs along all $\bR^{3,1}$ directions can be written as
\bea
\epsilon_+&=&\sigma_++ u \Gamma_+\Theta_-\sigma_-+(z\Gamma_z+x \Gamma_x) A \Theta_{+}\sigma_+~,~~~
\cr
\epsilon_-&=&\sigma_-+ r \Gamma_-\Theta_+ \sigma_++(z\Gamma_z+x \Gamma_x) A \Theta_{-}\sigma_-~.
\eea
This is derived using $\big(\Xi^{(\pm)}\big)^2\phi_\pm=0$ which in turn is implied from the integrability conditions (\ref{cc1}) and (\ref{cc3}) in the limit $\ell\rightarrow \infty$,
and $\Xi^{(\pm)}=A \Gamma_z\Theta_{\pm}$.

To verify (\ref{susyf}), it remains to count the multiplicity of solutions to (\ref{kseads41})  in the limit $\ell\rightarrow \infty$.
For this, if
$\sigma_-$ is a Killing spinor so is $\sigma_+= A^{-1} \Gamma_+\Gamma_z \sigma_-$ and vice versa if $\sigma_+$ is a Killing spinor so is $\sigma_-= A \Gamma_- \Gamma_z \sigma_+$.
In addition, it follows from direct observation that if $\sigma_+$ is a Killing spinor so is
\bea
\sigma_+'=\Gamma_{zx} \sigma_+~,
\eea
and similarly for the $\sigma_-$.  Therefore, {$\bR^{3,1}$  backgrounds preserve $4k$ supersymmetries confirming  (\ref{susyf}).

\subsubsection{$\bR^{n-1,1}$, $n>4$  backgrounds}

As in the previous cases, integrating the KSEs along the $\bR^{n-1,1}$ directions yields
\bea
\epsilon_+&=&\sigma_++ u \Gamma_+\Theta_-\sigma_-+\sum_a x^a\Gamma_a A \Theta_{+}\sigma_+~,~~~
\cr
\epsilon_-&=&\sigma_-+r \Gamma_-\Theta_+ \sigma_++\sum_a x^a\Gamma_a A \Theta_{-}\sigma_-~.
\eea
where $\Gamma_a$ are in the frame basis in $\bR^{n-1,1}$ and are transverse to the two lightcone directions $+$ and $-$, and $x^a$
are the corresponding coordinates. The dependence of the Killing spinors on the coordinates of $\bR^{n-1,1}$ depends
on $\sigma_\pm \notin \mathrm{Ker}\, \Xi^{(\pm)}$ even though $\big(\Xi^{(\pm)}\big)^2\sigma_\pm=0$ as a consequence of (\ref{cc1}) and (\ref{cc3}).
Observe that in the limit $\ell\rightarrow \infty$, $\Xi^{(\pm)}=A \Gamma_z \Theta_{\pm}$ and that $\Theta_{+}=\Theta_{-}$.

To verify (\ref{susyf}), it remains to count the multiplicity of solutions $\sigma_\pm$ of the remaining KSEs on $M^{11-n}$. As in the previous cases, if
$\sigma_-$ is a Killing spinor so is $\sigma_+= A^{-1} \Gamma_+\Gamma_z \sigma_-$ and vice versa if $\sigma_+$ is a Killing spinor so is $\sigma_-= A \Gamma_- \Gamma_z \sigma_+$.

Furthermore
for $\bR^{4,1}$ backgrounds it is easy to see that if $\sigma_+$ is a Killing spinor, then $\Gamma_{ab} \sigma_+$, $a<b$, is also a Killing spinor,
and similarly for the $\sigma_-$ spinors.  Therefore $\bR^{4,1}$  backgrounds preserve $8k$ supersymmetries confirming (\ref{susyf}).

Next consider the $\bR^{5,1}$.  In this case, if $\sigma_+$ is a Killing spinor, then $\Gamma_{ab}\sigma_+$, $a<b$ and $\Gamma_{a_1a_2a_3a_4}\sigma_+$, $a_1<\dots <a_4$,
are also Killing spinors. There are six $\Gamma_{ab}$, $a<b$ and a unique $\Gamma_{[4]}=\Gamma_{a_1a_2a_3a_4}$,  $a_1<\dots <a_4$, Clifford algebra operators. Moreover
$\Gamma_{[4]}$ commutes with all the KSEs (\ref{kseadsk1}) and (\ref{kseadsk2}).  Since $\Gamma_{[4]}^2=1$, $\sigma_+$ can lie in one of the two eigenspaces
of $\Gamma_{[4]}$.  If this is the case, then only three of the $\Gamma_{ab}\sigma_+$ are linearly independent. Thus   $\bR^{5,1}$ backgrounds
preserve $8k$ supersymmetries confirming (\ref{susyf}).

To  count the supersymmetries of $\bR^{6,1}$ backgrounds observe that if $\sigma_+$ is a Killing spinor so is $\Gamma_{ab}\sigma_+$, $a<b$ and $\Gamma_{a_1a_2a_3a_4}\sigma_+$, $a_1<\dots <a_4$.
There are ten $\Gamma_{ab}$, $a<b$ and five $\Gamma_{a_1a_2a_3a_4}\sigma_+$, $a_1<\dots <a_4$, Clifford algebra operators. Furthermore all $\Gamma_{a_1a_2a_3a_4}\sigma_+$, $a_1<\dots <a_4$
commute with the KSEs  (\ref{kseadsk1}) and (\ref{kseadsk2}).  Suppose we choose one such operator $\Gamma_{[4]}$.  Since $\Gamma_{[4]}^2=1$, $\Gamma_{[4]}$ has two eigenspaces
with eigenvalue $\pm1$. If $\sigma_+$ is in one of the two eigenspaces, then there are only eight linearly independent spinors $\Gamma_{ab}\sigma_+$, $a<b$ and $\Gamma_{a_1a_2a_3a_4}\sigma_+$, $a_1<\dots <a_4$. Thus $\bR^{6,1}$ backgrounds preserve $16k$ supersymmetries confirming (\ref{susyf}). Since there are no non-trivial $\bR^{6,1}$ backgrounds with 32 supersymmetries \cite{maxsusy},
such solutions necessarily preserve 16 supersymmetries.

\newsection{Concluding remarks}

We have systematically  described all warped AdS, $AdS_n\times_w M^{11-n}$, and flat backgrounds, $\bR^{n-1,1}\times_w M^{11-n}$, of D=11 supergravity which preserve at least one supersymmetry.
The novelty of our approach is that we have solved the KSEs of D=11 supergravity without making any assumptions
on the form of the fields and Killing spinors. Integrating over all $AdS_n$ and flat directions,  we have  identified all the a priori fractions of supersymmetry preserved by these backgrounds.  These results for AdS and flat
backgrounds have been summarized in equations (\ref{susyc}) and (\ref{susyf}) in the introduction, respectively.
Furthermore for AdS backgrounds that satisfy the requirements of maximum principle, we show that the Killing spinors
can be identified with the zero modes of Dirac like operators on $M^{11-n}$. This identification
is demonstrated via the proof of new Lichnerowicz type theorems for these Dirac like operators based on the identity (\ref{maxmaxmax}). As a consequence we show
that the number of Killing spinors of AdS backgrounds are given in terms of the dimension of the kernel of these Dirac like operators as in (\ref{susyg}).

The general solution of the KSE of 11-dimensional supergravity for $AdS_n\times M^{11-n}$ backgrounds has allowed us to investigate whether the Killing spinors
can be factorized as a product of a Killing spinor on $AdS_n$ and  a Killing spinor on the transverse $M^{11-n}$.  We have found that such a
factorization does not occur.  In particular we have demonstrated with an explicit example that assuming such a factorization gives
an incorrect counting for the supersymmetries of AdS backgrounds.

The identification of the a priori fractions of supersymmetry preserved by AdS and flat backgrounds can be used to find all such solutions of 11-dimensional supergravity and M-theory.
This  provides a systematic approach towards understanding all such backgrounds with applications in AdS/CFT and flux compactifications.
There are already some preliminary results in this direction. In particular we find that under certain assumptions that are no $AdS_6$
backgrounds in D=11 supergravity, however compare with \cite{passias} where the same results was proven assuming that the Killing spinors
factorize. We also find that the geometry of $AdS_2$ backgrounds is less restricted than
those investigated in \cite{kim}. The proof of these results will be presented elsewhere \cite{madsgeom}.

Another aspect of our approach is the generalization  of the classical Lichnerowicz  theorem to non-metric connections.
It is curious that the maximum principle which has been instrumental in the proof applies so widely in the
context of supergravity. It is not known why this is so and it should be investigated further, as ultimately it may be related to supersymmetry.

\vskip 0.5cm
\noindent{\bf Acknowledgements} \vskip 0.1cm
\noindent
JG is supported by the STFC grant, ST/1004874/1.
JG would like to thank the
Department of Mathematical Sciences, University of Liverpool for hospitality during which part of this work
was completed.
GP is partially supported by the  STFC rolling grant ST/J002798/1.
\vskip 0.5cm

\setcounter{section}{0}\setcounter{equation}{0}

\appendix{Notation and conventions}

Our form conventions are as follows. Let $\omega$ be a k-form, then
\bea
\omega={1\over k!} \omega_{i_1\dots i_k} dx^{i_1}\wedge\dots \wedge dx^{i_k}~,
\eea
and
\bea
d\omega={1\over k!} \partial_{i_1} \omega_{i_2\dots i_{k+1}} dx^{i_1}\wedge\dots \wedge dx^{i_{k+1}}~,
\eea
leading to
\bea
(d\omega)_{i_1\dots i_{k+1}}= (k+1) \partial_{[i_1} \omega_{i_2\dots i_{k+1}]}~.
\eea

Furthermore, we write
\bea
\omega^2= \omega_{i_1\dots i_k} \omega^{i_1\dots i_k}~,~~~\omega^2_{i_1 i_{2}}=\omega_{i_1j_1\dots j_{k-1}} \omega_{i_2}{}^{j_1\dots j_{k-1}} \ .
\eea
Given a volume form $d\mathrm{vol}={1\over n!} \epsilon_{i_1\dots i_n} dx^{i_1}\wedge \dots \wedge dx^{i_n}$, the Hodge dual of $\omega$ is defined as
\bea
\chi\wedge *\omega= (\chi, \omega) d\mathrm{vol}
\eea
where
\bea
(\chi, \omega)={1\over k!} \chi_{i_1\dots i_k} \omega^{i_1\dots i_k} \ .
\eea

It is well-known that for every form $\omega$, one can define a Clifford algebra element ${\slashed \omega}$ given by
\bea
{\slashed\omega}=\omega_{i_1\dots i_k} \Gamma^{i_1\dots i_k}
\eea
where $\Gamma^i$, $i=1,\dots n$, are the Dirac gamma matrices. In addition we introduce the notation
\bea
{\slashed\omega}_{i_1}= \omega_{i_1 i_2 \dots i_k} \Gamma^{i_2\dots i_k}~,~~~\slashed{\gom}_{i_1}= \Gamma_{i_1}{}^{
i_2\dots i_{k+1}} \omega_{i_2\dots i_{k+1}}~.
\eea
The rest of our spinor conventions can be found in \cite{Msystem}

\end{document}